\documentclass[]{article}

\usepackage[dvips]{epsfig}

\def\thebibliography#1{\section*{\normalsize \bf References 
 }\list
 {[\arabic{enumi}]}{\settowidth\labelwidth{[#1]}\leftmargin\labelwidth
 \advance\leftmargin\labelsep
 \usecounter{enumi}}
 \def\newblock{\hskip .11em plus .33em minus .07em}
 \sloppy\clubpenalty4000\widowpenalty4000
 \sfcode`\.=1000\relax}

\begin{document}

\sloppy

\twocolumn[

\begin{center} \large \bf 
  The moment sum rule and its consequences for ferromagnetism \\  
  in the Hubbard model
\end{center}
\vspace{-3mm}

\begin{center} 
  M. Potthoff, T. Herrmann, T. Wegner and W. Nolting
\end{center}
\vspace{-6mm}

\begin{center} \small \it 
   Lehrstuhl Festk\"orpertheorie,
   Institut f\"ur Physik, 
   Humboldt-Universit\"at zu Berlin, 
   Germany
\end{center}
\vspace{2mm}

\begin{center}
\parbox{141mm}{ 
The sum rule for the moments of the spectral density is discussed 
for the single-band Hubbard model. It is shown that respecting 
the sum rule up to the order $m=3$ is conceptually important for 
a qualitatively correct description of the quasi-particle band 
structure in the strong-correlation regime. Different analytical 
approximations for the self-energy are analyzed with respect to 
their compatibility with the moment sum rule. To estimate the 
practical usefulness of the sum rule, correlation functions and 
dynamical quantities are determined. The results obtained within 
the various approximation schemes of different complexity are 
compared with each other and also with essentially exact results 
available for infinite-dimensional lattices. It turns out that 
the $m=3$ moment is rather unimportant for the paramagnetic phase 
on the hyper-cubic lattice. Contrary, it decisively influences the 
magnetic phase boundary as well as the critical temperature for 
the ferromagnetic phase on an fcc-type lattice.
\vspace{4mm} 

}
\end{center}
\vspace{8mm} 
]

{\center \bf \noindent I. INTRODUCTION \\ \mbox{} \\} 

The Hubbard model \cite{Gut63,Hub63,Kan63} plays a central role 
in the attempts to understand the physics of correlated electrons 
on a lattice. It is surely oversimplified when it comes to a 
description of real materials such as the 3d transition metals 
and their oxides, for example. Nevertheless, it can provide deep
insight into the fundamental mechanisms that are responsible for 
various prominent correlation effects.

The Hubbard model sets up a notoriously difficult problem which 
now as before is not accessible to an exact solution in general.
An exception is given for the one-dimensional case ($d=1$) where 
a yet incomplete but very detailed understanding of the model 
properties in the whole parameter range has been achieved 
\cite{LW68,Eme79,Hal81,MS92,Voi95}. An important simplification 
of the model is also given in the opposite limit of high spatial 
dimensions $d=\infty$ \cite{MV89,MH89a,Vol93,GKKR96} which was 
recognized as a 
well-defined and non-trivial limiting case by Metzner and Vollhardt 
\cite{MV89}. Here the electronic self-energy is $\bf k$ independent 
or site-diagonal \cite{MH89b}, and thereby the model becomes 
equivalent to an effective impurity problem \cite{Jan91}.
An exact solution is possible by self-consistent mapping 
\cite{GK92a,Jar92} onto the single-impurity Anderson model (SIAM)
\cite{And61}, for example, followed by the numerical treatment 
\cite{HF86} within the Quantum-Monte-Carlo (QMC) approach 
\cite{Jar92,RZK92,GK92b,JP93}. In dimensions $d=3$ or $d=2$, 
however, one still has to resort to approximate treatments.

Valuable information that helps to judge of the reliability of a 
particular method is provided by exact identities, sum rules, 
limiting cases etc. In any dimension $d$ such rigorous results impose 
strong necessary conditions for the inevitable approximations. The 
purpose of the present paper is to focus on a particular sum rule: 
Exact expressions can be obtained for the (low-order) moments 
$\int E^m A_{{\bf k} \sigma}(E) \, dE$ of the spectral density 
$A_{{\bf k}\sigma}(E)$. We will argue that the first four moments 
($m=0-3$) yield valuable information on the quasi-particle band 
structure; they are especially important in the strong-coupling regime 
and also decisively influence the possibility and characteristics of 
spontaneous magnetic order. The moment sum rule has been considered 
not only within the context of the standard single-band Hubbard model 
in $d=\infty$ \cite{HN97a,PWN97,PHN97,WPN97},
in $d=3$ \cite{HL67,Nol72,NB89,NBdkB91,HN97b}, 
$d=2$ \cite{MEHM95,BE95,VT97} 
and $d=1$ \cite{MEHM95},
but also for the negative-$U$ case \cite{SPRN96} and for reduced
translational symmetry \cite{PN96,PN97b},
for the SIAM \cite{NO81}, 
for the $t$-$J$ \cite{Mas93} 
and for localized spin models \cite{CCD+84,BBR86,NO87} 
and may thus be of general interest for the construction of 
analytical approaches. 

There are several questions related to the moment sum rule which are 
not yet finally clarified. Firstly, we have to ask how to check 
to which order the sum rule is fulfilled for a particular (approximate)
method. Secondly, a kind of recipe is required that shows up how a 
given method can be modified to respect the sum rule up to a certain 
desired order. Thirdly, it should be worked out what is the actual 
conceptual improvement that is thereby achieved. Finally and most 
important, we need to know what can be achieved in practice, i.~e.\ 
whether the sum rule can help to come ``closer'' to the exact solution.

Especially the last question requires to compare with exact results.
As has been mentioned above, these are available for the 
Hubbard model in infinite dimensions. In the recent years there has 
been extensive work for $d=\infty$ lattices concerning the Fermi-liquid
phase at and off half-filling \cite{JP93}, the metal-insulator 
(Mott) transition \cite{Jar92,RZK92,GK92b,JP93,ZRK93,GK93,RKZ94}, 
transport properties \cite{PCJ93,JP94} and antiferromagnetic 
\cite{Jar92,JP93,ZRK93,GK93,RKZ94,FJ95} and ferromagnetic ordering 
\cite{JP93,VBH+97,Ulm98}. These QMC studies have been supplemented 
by the exact diagonalization method \cite{CK94,SRKR94} and by a 
number of approximate methods that are reliable in certain 
limits. Let us mention the non-crossing approximation (NCA) 
\cite{PCJ93,HS94,OPK97} as a strong-coupling approach, weak-coupling 
methods \cite{MH89c,MMH91,HC94,FLO+97}, variational approaches 
\cite{FMMH90,Uhr96,HUMH97} and others \cite{JV93,JMV93,LET96}. 
Moreover, 
the $d=\infty$ model is physically meaningful since the local 
approximation for the self-energy appears to be reasonable down 
to $d=3$ \cite{SC91} or even $d=2$ \cite{PJF95}. The essential 
physical properties of the $d=\infty$ Hubbard model are
thus expected to be comparable to those at $d=3$ ($d=2$).
For the present study we therefore restrict ourselves to the 
$d=\infty$ case where we are able to judge of the reliability 
of a particular method on the firm basis that is provided by available 
QMC results. This should help us to come to conclusive results
concerning the sum rule.

Our main idea is to elucidate the meaning and the usefulness of the 
moment sum rule by considering a number of standard approximations:
(i) the simple Hubbard-I approximation (H-I) \cite{Hub63},
(ii) Hubbard's alloy-analogy solution (AA) \cite{Hub64b},
(iii) the so-called Edwards-Hertz approach (EHA) \cite{EH}
in its improved version by Wermbter and Czycholl \cite{WC},
and finally (iv) the generalization \cite{MRetal} of the 
iterative perturbation theory \cite{GK92a} to arbitrary band-fillings
as proposed by Kajueter and Kotliar (KK) \cite{KK96}.
It turns out that all approaches are inconsistent with the moment 
sum rule for $m=3$ and yield the correct moments only up to $m=2$.

The moments of the spectral density are intimately related to the
coefficients in the high-energy expansion of the corresponding
self-energy \cite{Gor68}. We will show that thereby a possibility
is opened to improve upon a particular approximation analytically
and to ensure that the moment sum rule is respected up to $m=3$. 

This program has been completed successfully for the four approaches
mentioned above. The result is another set of methods, each evolving 
straightforwardly from its above counterpart:
(i) the spectral-density approach (SDA) \cite{Nol72,NB89,HN97a,HN97b},
(ii) the modified alloy-analogy (MAA) \cite{HN96,HN98},
(iii) the interpolating alloy-analogy-based approach (IAA) 
\cite{PHN97}, and
(iv) the modified perturbation theory (MPT) \cite{PWN97,WPN97}.

Each of these new approaches can be considered as a conceptual
improvement upon the original one: It can be shown that the 
additional inclusion of the fourth spectral moment does not 
affect their validity in different exactly solvable limiting cases. 
Furthermore, we will also show that significant improvement is 
achieved when comparing with the essentially exact QMC data of 
Refs.\ \cite{JP93,VBH+97,Ulm98}.

There is an interesting feature common to those four approaches that
yield the correct moments up to $m=3$: The analytical expression for 
the self-energy depends on a higher-order correlation function 
$B_\sigma$ which can be shown to be responsible for a (possibly) 
spin-dependent shift of the centers of gravity of the lower and 
the upper Hubbard band in the strong-correlation regime 
$U\mapsto \infty$. On the other hand, the original approaches 
are obtained if $B_\sigma$ is (ad hoc) replaced by its atomic-limit 
(H-I, AA) or by its Hartree-Fock value (EHA, KK), respectively.
One may thus expect that the inclusion of the fourth spectral
moment is especially important what concerns magnetic order.

After a general discussion of the moment sum rule and its conceptual
importance in the next section, we briefly discuss each of the 
methods in sections III-VI. Section VII presents new results for 
the paramagnetic phase on the hyper-cubic and for the ferromagnetic 
phase on an fcc-type lattice. For the discussion a comparative 
opposition of the different methods against each other and against 
QMC results appears to be helpful. Our intention is to arrive at 
general conclusions on the meaning and importance of the moment sum 
rule and of the correlation function $B_\sigma$ in particular which 
may be helpful for future analytical work.
\\

{\center \bf \noindent II. SPECTRAL MOMENTS AND $1/U$ PERTURBATION 
THEORY \\ \mbox{} \\} 

Let us start by recalling some essential facts for the limiting 
case of strong interaction ($U\gg t$). This will be important for 
the later discussion and for a concise formulation of the different 
approaches mentioned above. We first introduce the model as well as 
the basic physical quantities of interest. 

Using standard notations the Hubbard model reads:
\begin{equation}
  H = \sum_{ij\sigma} \left( T_{ij} - \mu \delta_{ij} \right)
  c^\dagger_{i\sigma} c_{j\sigma} 
  + \frac{1}{2} U \sum_{i \sigma} n_{i\sigma} n_{i -\sigma} \: .
\label{eq:hubbard}
\end{equation}
The hopping integrals $T_{ij}$ are assumed to be non-zero up to
nearest-neighbor distances: $T_{\langle ij \rangle} = - t$. The 
on-site energy $T_{ii} \equiv T_0$ defines the energy zero. We 
consider an infinite-dimensional lattice with the usual scaling
$d \, t^2 = \mbox{const.}$ for the hopping to retain the model
non-trivial \cite{MV89}. Restricting ourselves to homogeneous 
phases (para- and ferromagnetism), the one-electron Green function,
\begin{equation}
  G_{\sigma}(E) = \langle \langle c_{i\sigma} ; 
  c^\dagger_{i\sigma} \rangle \rangle_E \: ,
\label{eq:green}
\end{equation}
depends on the lattice geometry via the free Bloch density of states
(BDOS) only and can be written in the form:
\begin{equation}
  G_\sigma(E) = \int_{-\infty}^\infty \frac{\hbar 
  \rho^{\rm (B)}(z)}
  {E-(z-\mu) - \Sigma_{\sigma}(E)} \: dz \: ,
\label{eq:grhm}
\end{equation}
where the BDOS
\begin{equation}
  \rho^{\rm (B)}(E) = \frac{1}{N} \sum_{\bf k} 
  \delta(E-\epsilon({\bf k}))
\end{equation}
is given in terms of the tight-binding dispersion
\begin{equation}
  \epsilon({\bf k}) = \frac{1}{N} \sum_{ij} 
  e^{- i {\bf k} ({\bf R}_i-{\bf R}_j)} T_{ij} \: .
\end{equation}
The expression (\ref{eq:grhm}) for the Green function is based on
the fact that the self-energy $\Sigma_{\sigma}(E)$ is $\bf k$ 
independent or site-diagonal for $d=\infty$ \cite{MV89,MH89b}.
Let us also introduce the ${\bf k}$-resolved spectral density,
\begin{equation}
  A_{{\bf k}\sigma}(E) = -\frac{1}{\pi} \mbox{Im} \frac{\hbar}
     {E+i0^+ -(\epsilon({\bf k})-\mu) - \Sigma_\sigma(E+i0^+)} \: ,
\label{eq:kspden}
\end{equation}
and the on-site or $\bf k$-summed spectral density:
\begin{equation}
  A_{\sigma}(E) = \frac{1}{N} \sum_{\bf k} A_{{\bf k}\sigma}(E) \: .
\end{equation}

The atomic limit of vanishing hopping ($t=0$) represents the
zeroth-order result for the strong-coupling regime. We have:
\begin{equation}
  \Sigma_\sigma(E) = U n_{-\sigma} + 
  \frac{U^2 n_{-\sigma} (1 - n_{-\sigma}) }
  {E+\mu-T_0-U(1-n_{-\sigma})} \: ,
\label{eq:atlim}
\end{equation}
where $n_\sigma=\langle c^\dagger_{i\sigma} c_{i\sigma} \rangle$. 
Let us now consider the first non-trivial correction to the 
atomic solution (\ref{eq:atlim}) in the strong-coupling regime. 
For $U\mapsto \infty$ perturbational results can be derived 
by performing a canonical transformation of the Hubbard model 
\cite{HL67,CSO78,MGY88,EOMS94}. As has been shown first by 
Harris and Lange \cite{HL67}, the $1/U$ perturbation theory 
predicts the spectrum to be dominated by two charge-excitation 
peaks (Hubbard bands); the weight of additional satellites in the 
spectrum is of the order $(1/U)^4$ and can thus be neglected for 
strong interaction. Furthermore, at each $\bf k$ point in the 
Brillouin zone the center of gravity $T_{p\sigma}({\bf k})$ as 
well as the spectral weight $\alpha_{p\sigma}({\bf k})$ of the 
lower ($p=1$) and of the upper ($p=2$) Hubbard band can be 
calculated exactly. Specializing the results of Harris and 
Lange to the $d=\infty$ case, one obtains:
\begin{eqnarray}
  T_{1\sigma}({\bf k}) \!\!\! & = & \!\!\! 
  (1-n_{-\sigma}) \epsilon({\bf k}) + n_{-\sigma} B_{-\sigma}
  + {\cal O}(1/U) \: ,
  \nonumber \\
  T_{2\sigma}({\bf k}) \!\!\! & = & \!\!\! 
  U + n_{-\sigma} \epsilon({\bf k}) + (1-n_{-\sigma}) B_{-\sigma}
  + {\cal O}(1/U) \: ,
  \nonumber \\
  \alpha_{1\sigma}({\bf k}) \!\!\! & = & \!\!\! 
  1-n_{-\sigma} 
  \nonumber \\ 
  \!\!\! & + & \!\!\!
  \frac{2}{U} n_{-\sigma} (1-n_{-\sigma})
  ( B_{-\sigma} - \epsilon({\bf k}) ) 
  + {\cal O}(1/U)^2 \: ,
  \nonumber \\
  \alpha_{2\sigma}({\bf k}) \!\!\! & = & \!\!\! 
  1 - \alpha_{1\sigma}({\bf k}) \: ,
\label{eq:hl}
\end{eqnarray}
where the $B_{-\sigma}$ is defined as:
\begin{equation}
  {B}_{\sigma} = {T}_0 +
  \frac{1}{ n_{\sigma} ( 1 - n_{\sigma} ) }
  \sum_{j\ne i} {T}_{ij} 
  \langle c^\dagger_{i\sigma} c_{j\sigma} 
  (2 n_{i-\sigma} - 1) \rangle \: .
\label{eq:bdef}
\end{equation}
Contrary to the finite-dimensional case \cite{HN97b,MEHM95,BE95}, 
the $\bf k$ dependence of $T_{p\sigma}({\bf k})$ and 
$\alpha_{p\sigma}({\bf k})$ is exclusively due to the Bloch 
dispersion $\epsilon({\bf k})$. For finite $d$ an additional 
$\bf k$ dependence is introduced via $B_{-\sigma} \mapsto
B_{{\bf k}-\sigma}$.

From the result (\ref{eq:hl}) valuable ``global'' information on the 
quasi-particle band structure can be read off: Apart
from the Hubbard splitting, electron correlations manifest 
themselves in a specific narrowing of the lower and upper Hubbard 
band, in a redistribution of spectral weight among them and in a 
shift of their centers of gravity that is given by $n_{-\sigma} 
B_{-\sigma}$ and $(1-n_{-\sigma}) B_{-\sigma}$, respectively.
These (possibly spin-dependent) shifts, in particular, are expected 
to be important for ferromagnetic symmetry breaking: A spin-dependent
shift of the band center of gravity may generate and stabilize 
ferromagnetic solutions.

Here the question arises how it can be ensured that a given 
analytical approach is consistent with the exact perturbational
results of Harris and Lange for $U\mapsto \infty$. For this 
purpose we consider the moments of the spectral density which 
for $m=0,1,2,\dots$ are defined as:
\begin{equation}
  M^{(m)}_{{\bf k}\sigma} = \frac{1}{\hbar} \int_{-\infty}^\infty E^m
  A_{{\bf k}\sigma}(E) \, dE \: .
\label{eq:momentsdef}
\end{equation}
There is an alternative but equivalent way to calculate the moments:
\begin{equation}
  M^{(m)}_{{\bf k}\sigma} = \langle [ {\cal L}^m c_{{\bf k}\sigma} , 
  c_{{\bf k}\sigma}^\dagger ]_+ \rangle \: ,
\label{eq:moments}
\end{equation}
where ${\cal L O}=[{\cal O},H]_-$ denotes the commutator of an 
operator $\cal O$ with the Hamiltonian, and $[\cdots , \cdots]_+$ 
is the anticommutator. Eqs.\ (\ref{eq:momentsdef}) and 
(\ref{eq:moments}) represent the moment sum rule for the 
spectral density. In practice the sum rule is useful for low $m$ 
only. The limitation arises from the fact that with increasing $m$ 
Eq.\ (\ref{eq:moments}) involves equal-time correlation functions 
of higher and higher order which are usually unknown.

As has been shown in Ref.\ \cite{VT97}, the correctness of the
moments up to $m=2$ is a necessary condition for the evolution
of the Hubbard bands as $U\mapsto \infty$. On the other hand, it
is not sufficient as the following example shows: The self-consistent
second-order perturbation theory in the interaction $U$ for $d=\infty$
\cite{MH89c} predicts the correct moments up to $m=2$ but does not
yield the Hubbard splitting for large $U$ \cite{VT97,MH89c}.

To be consistent with the results of $1/U$ perturbation theory,
one has to check the existence of the Hubbard bands (which normally
can be done easily). Then, it is sufficient to ensure that the moment 
sum rule is fulfilled for $m=0-3$: For each ${\bf k}$ point of the 
Brillouin zone, the first four moments provide four pieces of 
information on the spectral density that (for $U\mapsto \infty$) 
unambiguously determine both, the centers of gravity and the weights, 
for the two Hubbard bands. Explicit expressions for the Hubbard model 
up to $m=3$ can be found Refs.\ \cite{NB89,PN96}, for example. The 
dispersions $T_{p\sigma}({\bf k})$ and weights 
$\alpha_{p\sigma}({\bf k})$ that result from the above reasoning
are the same as found by Harris and Lange in Eq.\ (\ref{eq:hl}).

In most analytical approaches one obtains an (approximate) 
expression for the self-energy. To make use of the sum rule, 
it becomes necessary to determine the spectral density from 
Eq.\ (\ref{eq:kspden}) and to calculate the moments by carrying
out the integration in Eq.\ (\ref{eq:momentsdef}). Finally, the 
(approximate) result for the moments has to be compared with the 
exact one given by Eq.\ (\ref{eq:moments}). 

The inconvenient integration in (\ref{eq:momentsdef}) can be avoided 
when considering the high-energy expansion of the Green function and 
of the self-energy: In the spectral representation of the on-site 
Green function,
\begin{equation}
  G_{\sigma}(E) = \int_{-\infty}^\infty \frac{A_{\sigma}(E')}
  {E-E'} \: dE' \: ,
\label{eq:gspden}
\end{equation}
we expand the denominator in powers of $1/E$. With Eq.\ 
(\ref{eq:momentsdef}) we immediately get:
\begin{equation}
  G_{\sigma}(E) = \hbar \sum_{m=0}^\infty 
  \frac{M_{\sigma}^{(m)}}{E^{m+1}} \: .
\label{eq:gexp}
\end{equation}
The coefficients in the high-energy expansion of the Green function 
are given by the (on-site) moments: $M_{\sigma}^{(m)} =
M_{ii\sigma}^{(m)} = N^{-1}\sum_{\bf k} M_{{\bf k}\sigma}^{(m)}$.
Via Eq.\ (\ref{eq:grhm}) they also 
determine the high-energy expansion coefficients of the self-energy:
\begin{equation}
  \Sigma_{\sigma}(E) = \sum_{m=0}^\infty 
  \frac{C^{(m)}_{\sigma}}{E^{m}} \: .
\label{eq:sexp}
\end{equation}
Using the explicit expressions for the moments as obtained from 
Eq.\ (\ref{eq:moments}) \cite{NB89,PN96}, a straightforward 
calculation yields:
\begin{eqnarray}
  C^{(0)}_{\sigma} \!\!\! & = & \!\!\! U n_{-\sigma} \: ,
  \nonumber \\
  C^{(1)}_{\sigma} \!\!\! & = & \!\!\! U^2 n_{-\sigma} 
  \left( 1 - n_{-\sigma} \right) \: ,
  \nonumber \\
  C^{(2)}_{\sigma} \!\!\! & = & \!\!\! U^2 n_{-\sigma}
  \left( 1 - n_{-\sigma} \right)
  \left( B_{-\sigma} - \mu
  + U (1- n_{-\sigma}) \right) \: .
  \nonumber \\
\label{eq:smom}
\end{eqnarray}
Any analytical (approximate) expression for the self-energy can 
easily be checked against these rigorous results simply by expanding 
in powers of $1/E$. Provided that the coefficients $C^{(m)}_{\sigma}$ 
turn out to be correct up to $m=2$, the moments of the resulting 
spectral density $M_{{\bf k}\sigma}^{(m)}$ will be correct up to 
$m=3$. For $U\mapsto \infty$ this ensures complete consistency with 
the $1/U$ perturbational results of Harris and Lange.

Finally, we have to express the expectation values in Eq.\ 
(\ref{eq:smom}) in terms of known quantities. For the 
(spin-dependent) average occupation number we have:
\begin{equation}
  n_\sigma \equiv \langle c^\dagger_{i\sigma} c_{i\sigma} \rangle
  = \frac{1}{2} + \frac{2}{\hbar \beta} \: \mbox{Re}
  \sum_{n=0}^\infty G_\sigma(iE_n) \: ,
\end{equation}
where $E_n = (2n+1) \pi / \beta$ and $\beta = 1 / k_{\rm B} T$. 
We also define the band-filling $n=n_\uparrow + n_\downarrow$.
Ferromagnetic order is indicated by an asymmetry $n_\uparrow \ne
n_\downarrow$ in the spin-dependent occupation numbers. In this
case the (dimensionless) spontaneous magnetization is given by 
$m=n_\uparrow - n_\downarrow$. We can also determine $B_\sigma$
(see Refs.\ \cite{PHN97,NB89,NBdkB91,PN96}, for example):
\begin{eqnarray}
  {B}_{\sigma} \!\!\!\! & = & \!\!\!\! {T}_{0} \, + \, 
  \frac{1}{n_{\sigma} ( 1 - n_{\sigma} )}
  \frac{2}{\hbar \beta} \, \mbox{Re} \sum_{n=0}^\infty 
  \left( \frac{2}{U} \Sigma_{\sigma}(iE_n) - 1 \right) 
  \nonumber \\ && \times
  \left[ 
  \left( iE_n - \Sigma_{\sigma}(iE_n) - T_0 + \mu \right) 
  G_{\sigma}(iE_n) - \hbar
  \right]
  \: . \nonumber \\
\label{eq:bfin}
\end{eqnarray}
Fortunately, it can be expressed in terms of the one-electron Green
function and self-energy, despite the fact that the definition of 
$B_\sigma$ (\ref{eq:bdef}) includes higher-order correlation functions.
\\

{\center \bf \noindent III. TWO-POLE GREEN FUNCTION \\ \mbox{} \\}

In the following we consider four different standard approximations
for the self-energy: The Hubbard-I solution (Sec.~III), the alloy 
analogy solution (Sec.~IV), the Edwards-Hertz approach (Sec.~V)
and the ansatz of Kajueter and Kotliar (Sec.~VI). Carrying out the 
high-energy expansion, we can check to which order $m$ the moment 
sum rule is fulfilled. Furthermore, we will show up how the
respective approaches can be corrected (if necessary) such that 
the sum rule is obeyed up to $m=3$.
\\

{\center \bf \noindent A. Hubbard-I approximation (H-I) \\ \mbox{} \\}

The so-called Hubbard-I solution is the simplest approach that
predicts a splitting of the non-interacting Bloch band into the 
two quasi-particle (Hubbard) subbands. In the original work 
\cite{Hub63} it is derived by decoupling the hierarchy of equations
of motion at the second level, i.~e.\ by assuming $\langle\langle 
c_{i\sigma} n_{i-\sigma} ; c^\dagger_{j\sigma} \rangle\rangle \approx 
\langle n_{i-\sigma} \rangle \langle\langle c_{i\sigma} ; 
c^\dagger_{j\sigma} \rangle \rangle$. From the resulting approximate
Green function and from the Dyson equation one obtains the Hubbard-I 
self-energy which is identical with the self-energy (\ref{eq:atlim})
of the atomic limit. The high-energy expansion is readily performed,
the expansion coefficients read:
\begin{eqnarray}
  C^{\rm (0,H-I)}_{\sigma} \!\!\! & = & \!\!\! U n_{-\sigma} \: ,
  \nonumber \\
  C^{\rm (1,H-I)}_{\sigma} \!\!\! & = & \!\!\! U^2 n_{-\sigma} 
  \left( 1 - n_{-\sigma} \right) \: ,
  \nonumber \\
  C^{\rm (2,H-I)}_{\sigma} \!\!\! & = & \!\!\! U^2 n_{-\sigma}
  \left( 1 - n_{-\sigma} \right)
  \left( T_0 - \mu
  + U (1- n_{-\sigma}) \right) \: .
  \nonumber \\
\label{eq:hiexp}
\end{eqnarray}
Comparing with (\ref{eq:smom}), we notice that the $m=2$ coefficient
(and thereby the $m=3$ moment) is incorrect.
\\

{\center \bf \noindent B. Spectral-density approach (SDA) \\ \mbox{} \\}

It is easily seen from the expansion coefficients that the sum rule
up to order $m=3$ can be restored if the atomic level $T_0$ in the H-I 
self-energy (\ref{eq:atlim}) is replaced by $T_0 \mapsto B_{-\sigma}$.
We obtain:
\begin{equation}
  \Sigma^{\rm (SDA)}_\sigma(E) = U n_{-\sigma} + 
  \frac{U^2 n_{-\sigma} (1 - n_{-\sigma}) }
  {E+\mu-B_{-\sigma}-U(1-n_{-\sigma})} \: .
\label{eq:sdasig}
\end{equation}
The Green function can be calculated from Dyson's equation, and 
from (\ref{eq:bfin}) we get $B_{-\sigma}$. We thus arrive at a 
conceptually simple improvement of the H-I solution that introduces 
an effective atomic level which is determined self-consistently. 

It turns out that this is identical with the spectral-density approach 
(SDA) \cite{NB89,BdKNB90,PN96,HN97b} in infinite dimensions. The SDA
is essentially equivalent to the Roth two-pole approximation for the 
Green function \cite{Rot69,BE95}. Furthermore, the SDA self-energy 
can also be obtained by means of the Mori-Zwanzig projection technique 
\cite{Mor65a,Zwa61,MEHM95}.
\\

{\center \bf \noindent IV. ALLOY ANALOGY \\ \mbox{} \\} 

The main idea of an alloy-analogy solution for the Hubbard model
is to consider the $-\sigma$ electrons to be ``frozen'' and to be 
randomly distributed over the sites of the lattice. The $\sigma$ 
electrons then move through a fictitious two-component alloy that 
is characterized by two atomic levels $E_{1\sigma}$ and $E_{2\sigma}$
and the concentrations $x_{1\sigma}$ and $x_{2\sigma}$. The
coherent potential approximation (CPA) \cite{VKE68} can be used 
to perform the configurational average over the positions of the 
frozen $-\sigma$ electrons. The self-energy for the $\sigma$ 
electrons is thus obtained via:
\begin{equation}
  0 = \sum_{p=1}^2 x_{p\sigma} \:
  \frac{E_{p\sigma} - \Sigma_\sigma(E) - T_0}
  {1 - \frac{1}{\hbar} G_\sigma(E) \:
  [E_{p\sigma} - \Sigma_\sigma(E) -T_0] } \: .
\label{eq:cpaeq}
\end{equation}
\\

{\center \bf \noindent A. Hubbard's alloy-analogy solution (AA) \\ \mbox{} \\}

Any two-component alloy analogy requires the specification of the two 
atomic levels $E_{p\sigma}$ and the concentrations $x_{p\sigma}$.
Within the conventional alloy-analogy solution due to Hubbard 
\cite{Hub64b} (AA) these are taken by referring to the atomic limit:
\begin{eqnarray}
  && E_{1\sigma}^{\rm (AA)} = T_0 \: , \hspace{8mm}
  E_{2\sigma}^{\rm (AA)} = T_0 + U \: ,
\nonumber \\ 
  && x_{1\sigma}^{\rm (AA)} = 1 - n_{-\sigma} \: , \hspace{8mm}
  x_{2\sigma}^{\rm (AA)} = n_{-\sigma} \: .
\label{eq:xaa}
\end{eqnarray}
Using this ansatz in the general CPA equation (\ref{eq:cpaeq}) and 
rearranging the terms, we arrive at:
\begin{equation}
  \Sigma^{\rm (AA)}_\sigma(E) = \frac{U n_{-\sigma}}{
  1 - \frac{1}{\hbar} G_\sigma(E) \:
  ( U - \Sigma^{\rm (AA)}_\sigma(E) )} \: .
\label{eq:cpa}
\end{equation}
This AA solution turns out to be exact for a special but non-trivial
limiting case of the Hubbard model: Switching off the hopping of 
the $-\sigma$ electrons only ($T_{ij} \mapsto T_{ij\sigma}$, 
$T_{ij-\sigma} = \delta_{ij} T_0$), defines the Falicov-Kimball 
model (FKM) \cite{FK69}. As has been shown by Brandt and Mielsch 
\cite{BM}, in infinite dimensions the exact self-energy
is given by Eq.\ (\ref{eq:cpa}). 

To get the high-energy expansion coefficients of the AA self-energy,
we insert the expansions (\ref{eq:gexp}) and (\ref{eq:sexp}) into the
CPA equation (\ref{eq:cpaeq}) and sort the different terms in powers 
of $1/E$. Considering all terms up to order $1/E^2$ yields the 
following set of equations:
\begin{eqnarray}
  1 \!\!\! & = & \!\!\!
  \sum_p x_{p\sigma} \: ,
  \nonumber \\
  0 \!\!\! & = & \!\!\!
  \sum_p x_{p\sigma} \: 
  (E_{p\sigma} - T_0 - C_\sigma^{(0)}) \: ,
  \nonumber \\
  0 \!\!\! & = & \!\!\! \sum_p x_{p\sigma} \:
  \Big[ (E_{p\sigma} - T_0 - C_\sigma^{(0)})^2 M_\sigma^{(0)}
  - C_\sigma^{(1)}
  \Big] \: ,
  \nonumber \\
  0 \!\!\! & = & \!\!\!
  \sum_p x_{p\sigma} \: 
  \Big[ (E_{p\sigma} - T_0 - C_\sigma^{(0)})^3 (M_\sigma^{(0)})^2
  \nonumber \\
  && + \;
  (E_{p\sigma} - T_0 - C_\sigma^{(0)})^2 M_\sigma^{(1)}
  \nonumber \\
  && - \;
  2 (E_{p\sigma} - T_0 - C_\sigma^{(0)})
  C_\sigma^{(1)} M_\sigma^{(0)} - C_\sigma^{(2)}
  \Big] \: .
\label{eq:cpahigh}
\end{eqnarray}
Inserting the atomic levels and concentrations from (\ref{eq:xaa})
as well as the exact low-order moments $M_\sigma^{(0)}=1$ and 
$M_\sigma^{(1)}=T_0-\mu+Un_{-\sigma}$ and solving for $C_\sigma^{(m)}$
results in:
\begin{equation}
  C_\sigma^{\rm (m, AA)} = C_\sigma^{(m)} 
  \Big|_{B_{-\sigma}\mapsto T_0} = C_\sigma^{\rm (m, H-I)} 
\end{equation}
for $m=0-2$. The high-energy expansion coefficients of the AA 
self-energy turn out to be identical to the coefficients within 
the H-I solution. Again the $m=2$ coefficient is found to be incorrect.
\\

{\center \bf \noindent B. Modified alloy-analogy (MAA) \\ \mbox{} \\} 

Simply replacing $T_0 \mapsto B_{-\sigma}$ in the CPA equation 
(\ref{eq:cpaeq}) and in (\ref{eq:xaa}) does not yield the correct
$m=2$ coefficient since $B_{-\sigma}$ would cancel out again in 
Eqs.\ (\ref{eq:cpahigh}). Another idea, however, turns out to 
be successful: The choice (\ref{eq:xaa}) for the levels and 
concentrations is rather obvious. On the other hand, it is by 
no means predetermined. We therefore consider $E_{p\sigma}$
and $x_{p\sigma}$ ($p=1,2$) as free parameters to be fixed
by the Eqs.\ (\ref{eq:cpahigh}) where $C_\sigma^{(m)}$ and 
$M_\sigma^{(m)}$ are taken to be the exact coefficients
(\ref{eq:smom}). This
automatically ensures the correctness of the moments up to 
$m=3$ and thereby provides an ``optimized'' alloy analogy.
Solving Eqs.\ (\ref{eq:cpahigh}) for the parameters of the
fictitious alloy yields:
\begin{eqnarray}
  E_{p\sigma}^{\rm (MAA)} \!\!\!\! & = & \!\!\!\! \frac{1}{2} 
  \left( T_0 + U + B_{-\sigma} \right) + (-1)^p \times
  \nonumber \\ && \hspace{-8mm}
  \sqrt{
  \frac{1}{4} \left( U + B_{-\sigma} - T_0 \right)^2 
  + U n_{-\sigma} \left( T_0 - B_{-\sigma} \right) 
  } 
  \nonumber \\ &&
\label{eq:maalevels}
\end{eqnarray}
and
\begin{equation}
  x_{1\sigma}^{\rm (MAA)} = 
  \frac{B_{-\sigma} + U (1 - n_{-\sigma}) - E_{1\sigma}^{\rm (MAA)}}
  {E_{2\sigma}^{\rm (MAA)} - E_{1\sigma}^{\rm (MAA)}}
  = 1 - x_{2\sigma}^{\rm (MAA)} \: .
\label{eq:maacons}
\end{equation}
It turns out that this result is identical to the result of the 
recently proposed modified alloy analogy (MAA) \cite{HN96,HN98}
where it was derived by referring to the split-band regime 
\cite{VKE68} of the CPA.

Inserting the result into the CPA equation (\ref{eq:cpaeq}) yields 
the MAA self-energy:
\begin{equation}
  \Sigma^{\rm {(MAA)}}_\sigma(E) = \frac{U n_{-\sigma}}{1 \; - \; 
  \frac{ \textstyle 
         \frac{1}{\hbar} G_\sigma(E)
         \left( U - \Sigma^{\rm (MAA)}_\sigma(E) \right)
       }
       { \textstyle 
         1 \; - \; \frac{1}{\hbar} G_\sigma(E)
         \left( B_{-\sigma} - T_0 \right)
       }
  } \: .
\label{eq:sigfinal}
\end{equation}
One recognizes that the MAA reduces to the AA if $B_{-\sigma}$ is 
replaced by $T_0$. As can be seen from Eq.\ (\ref{eq:bdef}) this 
is correct for the atomic limit and for the Falicov-Kimball model.
Just as the AA, the MAA therefore remains exact within these two 
limits. It also reduces to the AA for a paramagnet at half-filling
with symmetric BDOS where $B_{-\sigma} = T_0$ is 
required by particle-hole symmetry \cite{Mer77}.
\\

{\center \bf \noindent V. ALLOY-ANALOGY-BASED 
INTERPOLATION SCHEME \\ \mbox{} \\} 

A severe drawback of the alloy-analogy solutions consists in the 
fact that they are unable to reproduce the weak-coupling limit.
This defect can be eliminated if the CPA equation is considered
as a mere starting point for a reasonable interpolation formula
which is demanded to be exact for small $U$ as well as in the 
atomic limit and in the case of the FKM.

Standard perturbation theory in the interaction $U$ \cite{AGD64} 
provides us with the exact result for the self-energy in the 
weak-coupling regime. The first non-trivial perturbational result 
for the self-energy,
\begin{equation}
  \Sigma_{\sigma}(E) = U n_{-\sigma} + 
  \Sigma^{\rm (SOC)}_{\sigma}(E) \: ,
\label{eq:sigmaipt}
\end{equation}
beyond the Hartee term $U n_{-\sigma}$ is given by the second-order 
contribution (SOC):
\begin{eqnarray}
  && \hspace{-9mm} \Sigma^{\rm (SOC)}_{\sigma}(E) = 
  \frac{U^2}{\hbar^3} 
  \! \int \!\!\! \int \!\!\! \int 
  \frac{A^{(1)}_{\sigma}(x) A^{(1)}_{-\sigma}(y) 
  A^{(1)}_{-\sigma}(z)}{E-x+y-z} \times
  \nonumber \\ &&
  (f(x) f(-y) f(z) + f(-x) f(y) f(-z))
  \: dx \, dy \, dz \: .
  \nonumber \\ &&
\label{eq:sigma2}
\end{eqnarray}
Here $f(x)=1/(\exp(\beta x)+1)$ denotes the Fermi function, and 
$A^{(1)}_\sigma(E)$ is the free ($U=0$) spectral density being 
shifted in energy by a (possibly spin-dependent) constant
$E_\sigma$:
\begin{equation}
  A^{(1)}_{\sigma}(E) = A^{(0)}_{\sigma}(E-E_\sigma) \: ,
\label{eq:rho1}
\end{equation}
with $A^{(0)}_\sigma(E) = - \frac{1}{\pi} \mbox{Im} \, 
G_\sigma(E+i0^+)|_{U=0}$. The plain or conventional second-order
perturbation theory (SOPT) is recovered for $E_\sigma=0$, and with 
$E_\sigma=U n_{-\sigma}$ one obtains the SOPT around the Hartree-Fock 
solution (SOPT-HF) \cite{SC91,BJ90,PN97c}. 
To this end there is no need to specify the constant $E_\sigma$.
The expression (\ref{eq:sigma2}) is correct up to order $U^2$ for
any function $E_\sigma=E_\sigma(U)$ with $E_\sigma(0)=0$.
\\

{\center \bf \noindent A. Edwards-Hertz approach (EHA) \\ \mbox{} \\} 

An interpolation scheme that is based on the alloy-ana\-logy idea 
but correctly accounts for the weak-coup\-ling limit, has been 
suggested by Edwards and Hertz \cite{EH}. Within the Edwards-Hertz 
approximation (EHA) the self-energy has to be calculated from 
\cite{EH,WC}:
\begin{equation}
  \Sigma_\sigma^{\rm (EHA)}(E) = \frac{U n_{-\sigma}}{
  1 - \frac{1}{\hbar} \widetilde{G}_\sigma(E) \:
  ( U - \Sigma^{\rm (EHA)}_\sigma(E) )} \: .
\label{eq:sigeha}
\end{equation}
This differs from the conventional AA with respect to the propagator
$\widetilde{G}_\sigma(E)$: The Green function ${G}_\sigma(E)$ in 
Eq.\ (\ref{eq:cpa}) is (ad hoc) replaced by
\begin{equation}
  \widetilde{G}_\sigma(E) = 
  \frac{\hbar}{U^2 n_{-\sigma} (1-n_{-\sigma})} \:
  \Sigma^{\rm (SOC)}_\sigma
  (E-\Sigma^{\rm (EHA)}_\sigma(E)+E_\sigma) \: ,
\label{eq:mprop}
\end{equation}
where $\Sigma^{\rm (SOC)}_\sigma(E)$ is the second-order contribution 
to the SOPT self-energy introduced in Eq.\ (\ref{eq:sigma2}). A simple
calculation shows that the EHA reproduces the atomic and the FKM 
limit for arbitrary band-fillings provided that the shifts $E_\sigma$ 
in the definition (\ref{eq:rho1}) of the spectral density 
$A_\sigma^{(1)}(E)$ and in Eq.\ (\ref{eq:mprop}) are determined 
self-consistently from the condition:
\begin{equation}
  n_\sigma = \int_{-\infty}^\infty f(E) \, A_\sigma(E) \, dE 
  = \int_{-\infty}^\infty f(E) \, A^{(1)}_\sigma(E) \, dE \: .
\label{eq:nnhf}
\end{equation}
This improvement upon the original theory \cite{EH} has been 
introduced by Wermbter and Czycholl \cite{WC}. Expanding the
EHA self-energy in powers of $U$ up to order $U^2$, one recovers 
the exact weak-coupling result (\ref{eq:sigmaipt}).

The high-energy expansion is performed straightforwardly. For the
modified propagator one obtains:
\begin{equation}
  \frac{1}{\hbar} \widetilde{G}_\sigma(E) = 
  \frac{1}{E} +  
  \left( B^{\rm (HF)}_{-\sigma} - \mu + U n_{-\sigma} \right) 
  \frac{1}{E^2} + \cdots
  \: ,
\label{eq:gtilexp}
\end{equation}
where ${B}^{\rm (HF)}_{\sigma}$ is the Hartree-Fock decoupled 
$B_\sigma$ defined as:
\begin{equation}
  {B}^{\rm (HF)}_{\sigma} = {T}_0 +
  \frac{2n^{(1)}_{-\sigma}-1}{n^{(1)}_{\sigma}(1-n^{(1)}_{\sigma})}
  \sum_{j\ne i} {T}_{ij} 
  \langle c^\dagger_{i\sigma} c_{j\sigma} \rangle^{(1)}
\label{eq:b1def}
\end{equation}
and $\langle c^\dagger_{i\sigma} c_{j\sigma} \rangle^{(1)} =
-\frac{1}{\hbar \pi} \int dE f(E) \: \mbox{Im} \: 
G_{ij\sigma}(E+i0^+-E_\sigma)|_{U=0}$. With this result at hand,
the high-energy expansion coefficients are obtained from
Eq.\ (\ref{eq:sigeha}):
\begin{equation}
  C_\sigma^{\rm (m, EHA)} = C_\sigma^{(m)} 
  \Big|_{B_{-\sigma} \mapsto B^{\rm (HF)}_{-\sigma}} \: . 
\label{eq:ehahigh}
\end{equation}
Again, the first two coefficients $C^{(0)}_{\sigma}$ and 
$C^{(1)}_{\sigma}$ are predicted correctly while the $m=2$
coefficient turns out to be wrong. Only in the atomic and
in the FKM limit as well as for the symmetric case $n=1$ we
have $B_{-\sigma} = {B}^{\rm (HF)}_{-\sigma} = T_0$.
\\

{\center \bf \noindent B. Interpolating 
alloy-analogy approach (IAA) \\ \mbox{} \\} 

The EHA interpolation formula can be improved such that the moments
up to $m=3$ are reproduced correctly: The interpolating alloy-analogy 
approach (IAA) \cite{PHN97} starts from the CPA equation 
(\ref{eq:cpaeq}) with the Green function replaced by the modified 
propagator $\widetilde{G}_\sigma(E)$ which is given by Eq.\
(\ref{eq:mprop}):
\begin{equation}
  0 = \sum_{p=1}^2 x_{p\sigma}^{\rm (IAA)} \:
  \frac{E_{p\sigma}^{\rm (IAA)} - \Sigma_\sigma(E) - T_0}
  {1 - \frac{1}{\hbar} \widetilde{G}_\sigma(E) \:
  [E_{p\sigma}^{\rm (IAA)} - \Sigma_\sigma(E) -T_0] } \: .
\label{eq:iaacpaeq}
\end{equation}
The high-energy expansion of this equation is then used to determine 
the (unknown) atomic levels and concentrations. The only difference
with respect to the expansion (\ref{eq:cpahigh}) of the general CPA 
equation is that the coefficients $M_\sigma^{(m)}$ for $m=0,1$ have 
now to be interpreted as the moments of the modified propagator. For 
$m=0,1$ they can be read off from Eq.\ (\ref{eq:gtilexp}). Solving
for the unknowns yields the following result:
\begin{eqnarray}
  E_{p\sigma}^{\rm (IAA)} \!\!\!\! & = & \!\!\!\! T_0 + \frac{1}{2} 
  \left( U + B_{-\sigma} - B^{\rm (HF)}_{-\sigma} \right) 
  + (-1)^p \times
  \nonumber \\ && \hspace{-14mm}
  \sqrt{
  \frac{1}{4} \left( U + B_{-\sigma} - B^{\rm (HF)}_{-\sigma} \right)^2 
  + U n_{-\sigma} \left( B^{\rm (HF)}_{-\sigma} - B_{-\sigma} \right) 
  } \: ,
  \nonumber \\ 
  x_{1\sigma}^{\rm (IAA)} \!\!\! & = & \!\!\!
  \frac{B_{-\sigma} - B^{\rm (HF)}_{-\sigma} + T_0
  + U (1 - n_{-\sigma}) - E_{1\sigma}^{\rm (IAA)}}
  {E_{2\sigma}^{\rm (IAA)} - E_{1\sigma}^{\rm (IAA)}} 
  \nonumber \\ 
  \!\!\! & = & \!\!\! 1 - x_{2\sigma}^{\rm (IAA)} \: .
\label{eq:iaacons}
\end{eqnarray}
Inserting into (\ref{eq:iaacpaeq}) results in:
\begin{equation}
  \Sigma^{\rm {(IAA)}}_\sigma(E) = \frac{U n_{-\sigma}}{1 \; - \; 
  \frac{ \textstyle 
         \frac{1}{\hbar} \widetilde{G}_\sigma(E)
         \left( U - \Sigma^{\rm (IAA)}_\sigma(E) \right)
       }
       { \textstyle 
         1 \; - \; \frac{1}{\hbar} \widetilde{G}_\sigma(E)
         \left( B_{-\sigma} - B_{-\sigma}^{(1)} \right)
       }
  } \: .
\label{eq:sigiaa}
\end{equation}
This differs from the MAA self-energy with respect to the modified 
propagator; also $T_0$ is replaced by $B_{-\sigma}^{\rm (HF)}$
in (\ref{eq:sigiaa}). By construction the moments of the spectral
density are correct up to $m=3$ in the IAA. Furthermore, the 
theory is exact up to order $U^2$ and in the atomic and FKM limit.
\\

{\center \bf \noindent VI. ITERATIVE PERTURBATION THEORY \\ \mbox{} \\} 

While for the alloy-analogy-based theories the correct weak-coupling 
behavior has to be enforced artificially, it can be automatically 
accounted for within a diagrammatic approach. Furthermore, a 
diagrammatic ansatz appears to be attractive since it allows 
to construct thermodynamically consistent theories which are 
conserving in the sense of Kadanoff and Baym \cite{BK61} and 
thus respect Luttinger's sum rule \cite{LW60}. The simplest 
conserving approximation is given by Hartree-Fock theory.
It recovers the $m=0$ and $m=1$ moments only. One additional 
moment ($m=2$) is correct within the self-consistent second-order 
perturbation theory (SC-SOPT) \cite{MH89c,HC94}. No improvement, 
however, is achieved by higher-order conserving approximations: On
the contrary, within the fluctuation-exchange approximation (FLEX) 
\cite{BSW89} and also for its analogue in $d=\infty$ \cite{MMH91}, 
already the $1/E$ coefficient in the expansion of the self-energy 
and thus the $m=2$ moment turns out to be incorrect.

The (non-conserving) SOPT around the Hartree-Fock solution (SOPT-HF) 
\cite{SC91,BJ90,PN97c} yields the correct moments up to $m=1$.
For a paramagnet at half-filling, however, the moments are correct 
even up to $m=3$ (in infinite dimensions). Furthermore, for $t=0$ 
SOPT-HF recovers the atomic-limit solution (\ref{eq:atlim}) and may
thus provide a reasonable interpolation between the weak- and the 
strong-coupling regime at $n=1$.

In infinite dimensions the Hubbard model can be self-consistently
mapped onto the single-impurity Anderson model (SIAM) 
\cite{GK92a,Jar92}:
\begin{eqnarray}
  H_{\rm SIAM} \!\!\!\! & = & \!\!\!\!
  \sum_\sigma (\epsilon_d - \mu)
  c_{\sigma}^\dagger c_{\sigma}
  + \sum_{k\sigma} (\epsilon_{k\sigma} - \mu)
  a_{k\sigma}^\dagger a_{k\sigma} 
  \nonumber \\ && \hspace{-10mm} 
  + \:
  U n_{\sigma} n_{-\sigma} +
  \sum_{k\sigma} V_{k\sigma} \left(
  c_{\sigma}^\dagger a_{k\sigma} +
  a_{k\sigma}^\dagger c_{\sigma}
  \right) \: .
\end{eqnarray}
The on-site Green function and the local self-energy of the $d=\infty$
Hubbard model are identical with the impurity Green function and
self-energy provided that the conduction band dispersion 
$\epsilon_{k\sigma}$ and the hybridization strength $V_{k\sigma}$ 
in the SIAM are chosen such that the self-consistency condition,
\begin{equation}
  \Delta_\sigma(E+\mu) = E - (\epsilon_d - \mu) - \Sigma_{\sigma}(E)
  - \hbar \left( G_{\sigma}(E) \right)^{-1} \: ,
\label{eq:sc}
\end{equation}
for the hybridization function $\Delta_\sigma(E)=\sum_k V^2_{k\sigma} /
(E-\epsilon_{k\sigma})$ is fulfilled. 

Within the context of the {\em symmetric} single-impurity Anderson 
model (SIAM), SOPT-HF has 
been recognized to be quite well behaved \cite{YYSZHS}. 
Furthermore, as for the Hubbard model, SOPT-HF respects the moment sum 
rule up to $m=3$ and the atomic limit of vanishing hybridization. The 
so-called iterative perturbation theory (IPT) \cite{GK92a} 
makes use of these advantages by combining SOPT-HF for the SIAM with 
the self-con\-sis\-tent 
mapping of the $d=\infty$ Hubbard model onto the SIAM. Indeed, IPT 
yields convincing results as has been proven by comparison with QMC 
studies \cite{ZRK93,GK93,RKZ94}. The advantageous properties of the
IPT are lost, however, if one considers the non-symmetric case 
$n\ne 1$.
\\

{\center \bf \noindent A. Kajueter-Kotliar approach (KK) \\ \mbox{} \\} 

Kajueter and Kotliar \cite{KK96} proposed an interpolating ansatz for 
the self-energy of the SIAM which for arbitrary fillings is correct 
in the weak-coupling and the atomic limit. For the symmetric case the 
Kajueter-Kotliar approach (KK) reduces to the usual IPT. 
The interpolative self-energy is given by:
\begin{equation}
  \Sigma_{\sigma}(E) = U n_{-\sigma} +
  \frac{a_\sigma \Sigma^{\rm (SOC)}_{\sigma}(E)}
  {1 - b_\sigma \Sigma^{\rm (SOC)}_{\sigma}(E)} \: ,
\label{eq:ansatz}
\end{equation}
where $n_{\sigma}=\langle c^\dagger_\sigma c_\sigma \rangle$ is the 
average occupancy of the impurity site and 
$\Sigma^{\rm (SOC)}_{\sigma}(E)$ 
the second-order contribution within SOPT-HF for the SIAM.
The coefficients $a_\sigma$ and $b_\sigma$ as well as a fictitious
chemical potential $\widetilde{\mu}_\sigma$ that appears in the 
definition of the Hartree-Fock spectral density
[cf.\ Eq.\ (\ref{eq:sigma2})],
\begin{equation}
  A^{(1)}_{\sigma}(E) = - \frac{1}{\pi} \mbox{Im} \frac{\hbar} 
  {E + \widetilde{\mu}_\sigma - \epsilon_d
  - \Delta_\sigma(E+\mu) - U n_{-\sigma}} \: ,
\label{eq:g1}
\end{equation}
are considered as free parameters. $\widetilde{\mu}_\sigma$ is fixed
by requiring $\mu=\mu|_{U=0}+\Sigma(0)$, i.~e.\ the Luttinger sum
rule in $d=\infty$ \cite{MH89c}. For any choice of $a_\sigma$ and
$b_\sigma$ the first two moments $m=0$ and $m=1$ are correct. 
Within the KK approach, the parameters $a_\sigma$ and $b_\sigma$
are determined from the moment sum rule for $m=2$ and the exact
atomic-limit solution \cite{KK96}. This yields:
\begin{equation}
  a_\sigma = \frac{n_{-\sigma} 
  (1 - n_{-\sigma})}
  {n_{-\sigma}^{(1)} 
  ( 1 - n_{-\sigma}^{(1)} )}
\label{eq:apar}
\end{equation}
and 
\begin{equation}
  b_\sigma = \frac{U (1-2n_{-\sigma}) + \widetilde{\mu}_\sigma-\mu}
  {U^2 n_{-\sigma}^{(1)} (1-n_{-\sigma}^{(1)})} \: ,
\label{eq:bpar}
\end{equation}
where $n_{\sigma}^{(1)} = \hbar^{-1} \int f(E) A^{(1)}_{\sigma}(E) dE$.

This procedure, however, leads to an incorrect result for the $m=3$
moment. Inserting (\ref{eq:apar}) and (\ref{eq:bpar}) into the ansatz
(\ref{eq:ansatz}), expanding in powers of $1/E$, and comparing with
the exact result (\ref{eq:smom}), we have
\begin{equation}
  C_\sigma^{\rm (m, KK)} = C_\sigma^{(m)} 
  \Big|_{B_{-\sigma} \mapsto B^{\rm (HF)}_{-\sigma}} 
\label{eq:kkhigh}
\end{equation}
up to $m=2$. Here, within the context of the SIAM we have introduced
the following definitions:
\begin{equation}
  {B}_{\sigma} = \epsilon_d +
  \frac{1}{ n_{\sigma} ( 1 - n_{\sigma} ) }
  \sum_{k} {V}_{k\sigma} 
  \langle a^\dagger_{k\sigma} c_{\sigma} 
  (2 n_{-\sigma} - 1) \rangle 
\label{eq:bsiam}
\end{equation}
and
\begin{equation}
  B_{\sigma}^{\rm (HF)} = \epsilon_d + 
  \frac{2n_{-\sigma}^{(1)}-1}{n_{\sigma}^{(1)} 
  (1-n_{\sigma}^{(1)})} \sum_{k} V_{k\sigma}
  \langle a_{k\sigma}^\dagger c_{\sigma} \rangle^{(1)} 
\label{eq:b1siam}
\end{equation}
which are related to the corresponding definitions for the Hubbard 
model (\ref{eq:bdef}) and (\ref{eq:b1def}) by means of the 
self-consistent mapping. As for the EHA the moment sum rule for $m=3$
is violated since generally $B^{\rm (HF)}_{-\sigma} \ne B_{-\sigma}$. 
\\

{\center \bf \noindent B. Modified 
perturbation theory (MPT) \\ \mbox{} \\} 

It is possible to improve the Kajueter-Kotliar approach such that
the moments up to $m=3$ are recovered correctly \cite{PWN97}. This 
constitutes the modified perturbation theory (MPT) \cite{WPN97}. We 
start from the same ansatz (\ref{eq:ansatz}) for the self-energy 
of the SIAM. Contrary to the KK approach, however, the parameters
are fitted to the $m=2$ and the $m=3$ moment. Using 
\begin{eqnarray}
  && \mbox{} \hspace{-8mm} 
  \Sigma_{d\sigma}^{\rm (SOC)}(E) = 
  \frac{U^2n^{(1)}_{-\sigma}(1-n^{(1)}_{-\sigma})}{E} 
  \nonumber \\ && 
  + \frac{U^2n^{(1)}_{-\sigma}(1-n^{(1)}_{-\sigma})
  (B_{-\sigma}^{\rm (HF)}-\widetilde{\mu}_\sigma+Un_{-\sigma})}
  {E^2} + \cdots \: , \nonumber \\
\end{eqnarray}
the high-energy expansion of the ansatz (\ref{eq:ansatz}) is easily
performed. Comparing with the exact result for the expansion 
coefficients in Eq.\ (\ref{eq:smom}), we have to choose
\begin{equation}
  b_\sigma = \frac{ B_{-\sigma} - \mu - B^{\rm (HF)}_{-\sigma} 
  + \widetilde{\mu}_\sigma 
  + U (1 - 2 n_{-\sigma}) ) }
  {U^2 n_{-\sigma}^{(1)} 
  ( 1 - n_{-\sigma}^{(1)} )} 
\label{eq:bparnew}
\end{equation}
for the parameter $b_\sigma$ to ensure the correctness of the moments.
$a_\sigma$ is still given by Eq.\ (\ref{eq:apar}).

We also modify the condition that fixes the parameter 
$\widetilde{\mu}_\sigma$. Analogous to the condition (\ref{eq:nnhf})
used in the EHA and IAA, we demand:
\begin{equation}
  n_{\sigma}^{(1)} = 
  n_{\sigma} \: .
\label{eq:c2}
\end{equation}
Choosing $\widetilde{\mu}_\sigma$ to enforce the Luttinger theorem
implies the theory to be intrinsically limited to zero temperature.
The condition (\ref{eq:c2}) is less problematic. It also allows 
to perform calculations at finite temperatures. Numerical results 
for $T=0$ have shown \cite{WPN97} that a significant violation of the
Luttinger sum rule occurs for fillings that considerably differ 
from half-filling only.

The MPT reduces to the KK approach for the symmetric case of a 
paramagnet at half-filling and symmetric BDOS: Here we
have $B^{\rm (HF)}_{\sigma} = B_{\sigma} = \epsilon_d$ due to
particle-hole symmetry. For small $U$ the MPT recovers the exact
weak-coupling result up to order $U^2$. By construction the 
moment sum rule is obeyed up to $m=3$ which ensures the theory 
to be consistent with the strong-coupling perturbational results 
of Harris and Lange and also implies the validity of the MPT in 
the atomic limit for arbitrary filling.

For the following discussion we refer to the KK approach as the
original approach of Ref.\ \cite{KK96} but with the Luttinger
theorem replaced by Eq.\ (\ref{eq:c2}) as the condition to fix
$\widetilde{\mu}_\sigma$. This allows to compare the ``KK''
approach with the MPT also for $T\ne 0$.
\\

{\center \bf \noindent VII. RESULTS AND DISCUSSION \\ \mbox{} \\} 

Numerical calculations have been performed for all approaches that
have been discussed in the preceding sections. To study the effect
of $B_\sigma$ and the meaning of the moment sum rule for the
paramagnet, we consider the hyper-cubic (hc) lattice in infinite
dimensions. The coordination number is $Z=2d$. We take
$T_{\langle ij \rangle} \equiv t = t^\ast/\sqrt{2Z}$ 
($t^\ast=\mbox{const.}$) for the
scaling of the hopping as $d\mapsto \infty$ \cite{MV89}. The 
non-interacting Bloch density of states (BDOS) is a Gaussian
\cite{MV89}:
\begin{equation}
  \rho^{\rm (B)}(E) = \frac{1}{t^\ast \sqrt{\pi}} 
  e^{ -\left( E/t^\ast \right)^2 } \: .
\label{eq:bdoshc}
\end{equation}
$T_0 \equiv T_{ii}=0$ has been chosen to fix the energy zero.

We also consider a generalization of the $d=3$ fcc lattice to 
infinite dimensions \cite{MH91} which favors ferromagnetic order
\cite{VBH+97,Ulm98}. The hopping is scaled as 
$t = t^\ast/\sqrt{Z}$ where $Z=2d(d-1)$, and the BDOS reads:
\begin{equation}
  \rho^{\rm (B)}(E) = \frac{e^{-(1+\sqrt{2}E/t^\ast)/2}}
  {t^\ast \sqrt{\pi (1+\sqrt{2}E/t^\ast)}}
  \: .
\label{eq:bdosfcc}
\end{equation}
Energy units are chosen such that $t^\ast = 1$.
\\

{\center \bf \noindent A. Paramagnet \\ \mbox{} \\} 

Let us consider the hc lattice first. In a wide region of the
$U$-$n$ plane the system is a paramagnetic Fermi liquid. Near
half-filling antiferromagnetic order is observed
\cite{Jar92,JP93,ZRK93,GK93,RKZ94,FJ95}. Saturated ferromagnetism
can be excluded completely \cite{Uhr96} while non-saturated 
ferromagnetic order has been found recently for very strong 
interaction $U$ \cite{OPK97}.
We will restrict our investigation for the hc lattice to the 
paramagnetic phase only.

According to the results of Harris and Lange, the first non-trivial
effect of $1/U$ perturbation theory beyond the atomic limit consists
in a specific shift of the centers of gravity of the Hubbard bands.
For the lower one it is given by $n_{-\sigma} B_{-\sigma}$. Fig.~1
shows this band shift as a function of the filling $n$ for the hc
lattice at finite inverse temperature $\beta = 7.2$ and at moderate
coupling strength $U=4$ (the $U$ dependence will be discussed below).
For each of the approaches that obey the moment sum rule up to 
$m=3$, the correlation function $B_{-\sigma}$ (inset in Fig.~1) has
been determined self-consistently via Eq.\ (\ref{eq:bfin}).
Electron correlations are meaningless for filling $n=0$;
consequently, the band shift $n_{-\sigma} B_{-\sigma}$ vanishes. 
At half-filling $n=1$ we have $n_{-\sigma} B_{-\sigma} = 0$ due to
particle-hole symmetry. Inbetween the band shift acquires positive
values with a maximum at $n \approx 0.5 - 0.75$ depending on the 
approximation scheme considered. The typical size of the shift
($n_{-\sigma} B_{-\sigma} \approx 0.2$) is significant compared
with the variance $\Delta = 1/\sqrt{2}$ of the BDOS.

\begin{figure}[b]
\vspace{-8mm}
\centerline{\psfig{figure=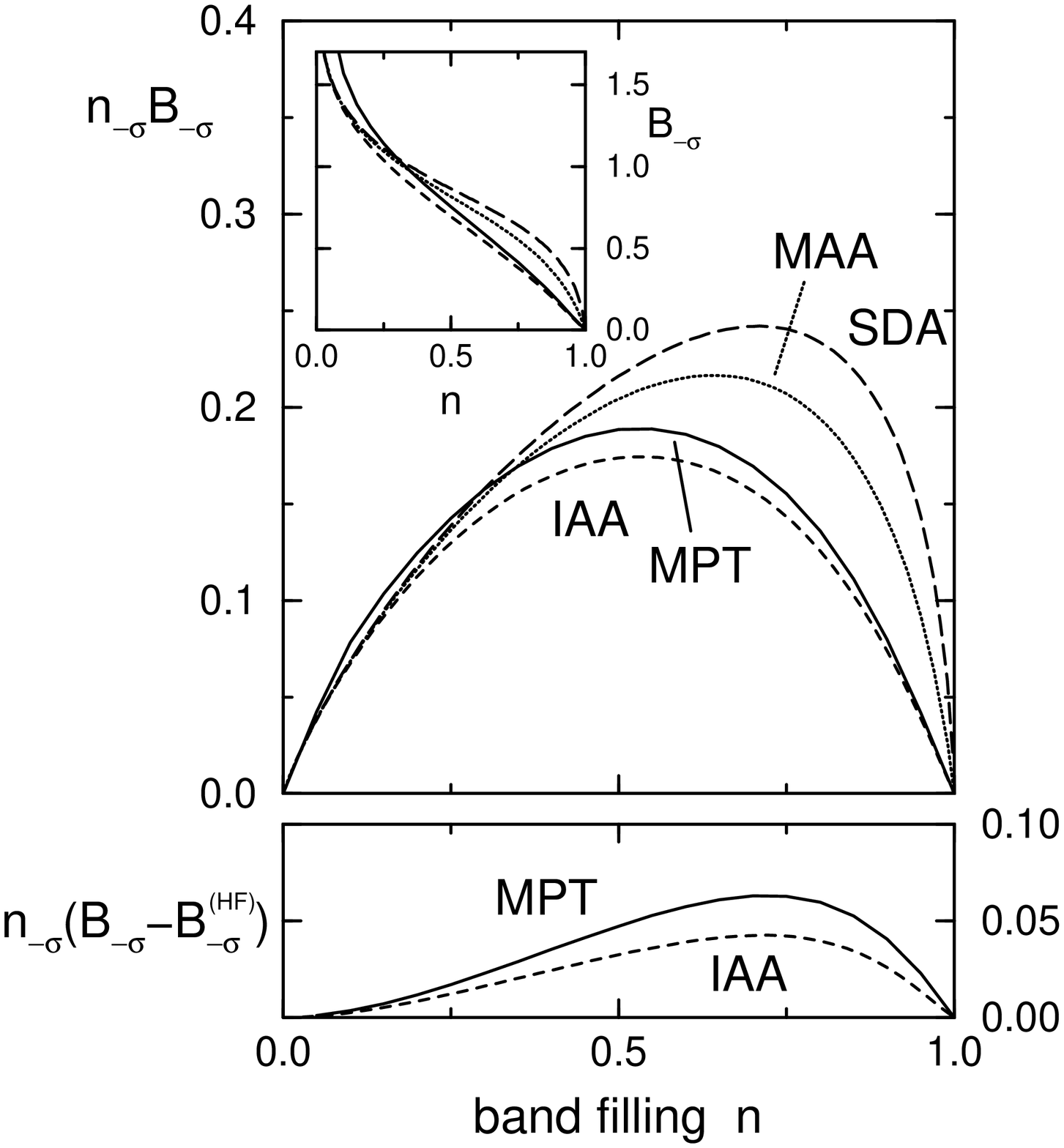,width=75mm,angle=0}}
\vspace{0mm}
\parbox[]{85mm}{\small Fig.~1.
Filling dependence of the shift $n_{-\sigma} B_{-\sigma}$ of the 
center of gravity of the lower Hubbard band as obtained 
self-consistently within the spectral-density approach (SDA), the 
modified alloy analogy (MAA), the interpolating alloy-analogy-based
approach (IAA) and the modified perturbation theory (MPT).
Inset: filling-dependence of $B_{-\sigma}$. Lower panel:
difference $n_{-\sigma} (B_{-\sigma} - B_{-\sigma}^{\rm (HF)})$
for the IAA and MPT. Calculations for the $d=\infty$ paramagnetic
Hubbard model on the hc lattice at $U=4$ and $\beta=7.2$. The
results are symmetric to the $n=1$ axis (half-filling). Energy
units are chosen such that $t^\ast = 1$.
}
\end{figure}

For the same set of parameters but with the filling fixed at
$n=0.79$, Fig.~2 shows the quasi-particle density of states (DOS)
$A_\sigma(E)$. For the discussion of the effects of the band shift
we concentrate on the result for the spectral-density approach (SDA)
and the Hubbard-I solution (H-I) first. Within the two-pole
approaches the DOS for $U \mapsto \infty$ can be written in terms
of the BDOS as:
\begin{eqnarray}
  A_\sigma(E) \!\!\!\! &=& \!\!\!\!  
  \rho^{\rm (B)}\left( \frac{E+\mu-n_{-\sigma} B_{-\sigma}}
  {1-n_{-\sigma}} \right) 
  \nonumber \\ \!\!\!\! &+& \!\!\!\! 
  \rho^{\rm (B)}\left( \frac{E+\mu-U-(1-n_{-\sigma}) B_{-\sigma}}
  {n_{-\sigma}} \right) \: ,
\label{eq:qdos2p}
\end{eqnarray}
where $B_{-\sigma} \mapsto T_0$ for the H-I solution.
Comparing the SDA with the H-I solution, the shift 
$n_{-\sigma} B_{-\sigma}$ of the lower Hubbard band present in the
SDA must be canceled exactly by a corresponding shift of the 
chemical potential $\mu$ for a fixed filling $n$. Thus the small
differences between SDA and H-I seen in the lower Hubbard band
exclusively result from the finite value for $U$ and will disappear
for $U\mapsto \infty$. On the other hand, the upper Hubbard band 
is significantly shifted to higher energies compared with the H-I
solution. This effect persists also for $U\mapsto \infty$ where
the energetic positions will differ by an amount 
$(1-2n_{-\sigma}) B_{-\sigma}$.

\begin{figure}[b]
\vspace{7mm}
\centerline{\psfig{figure=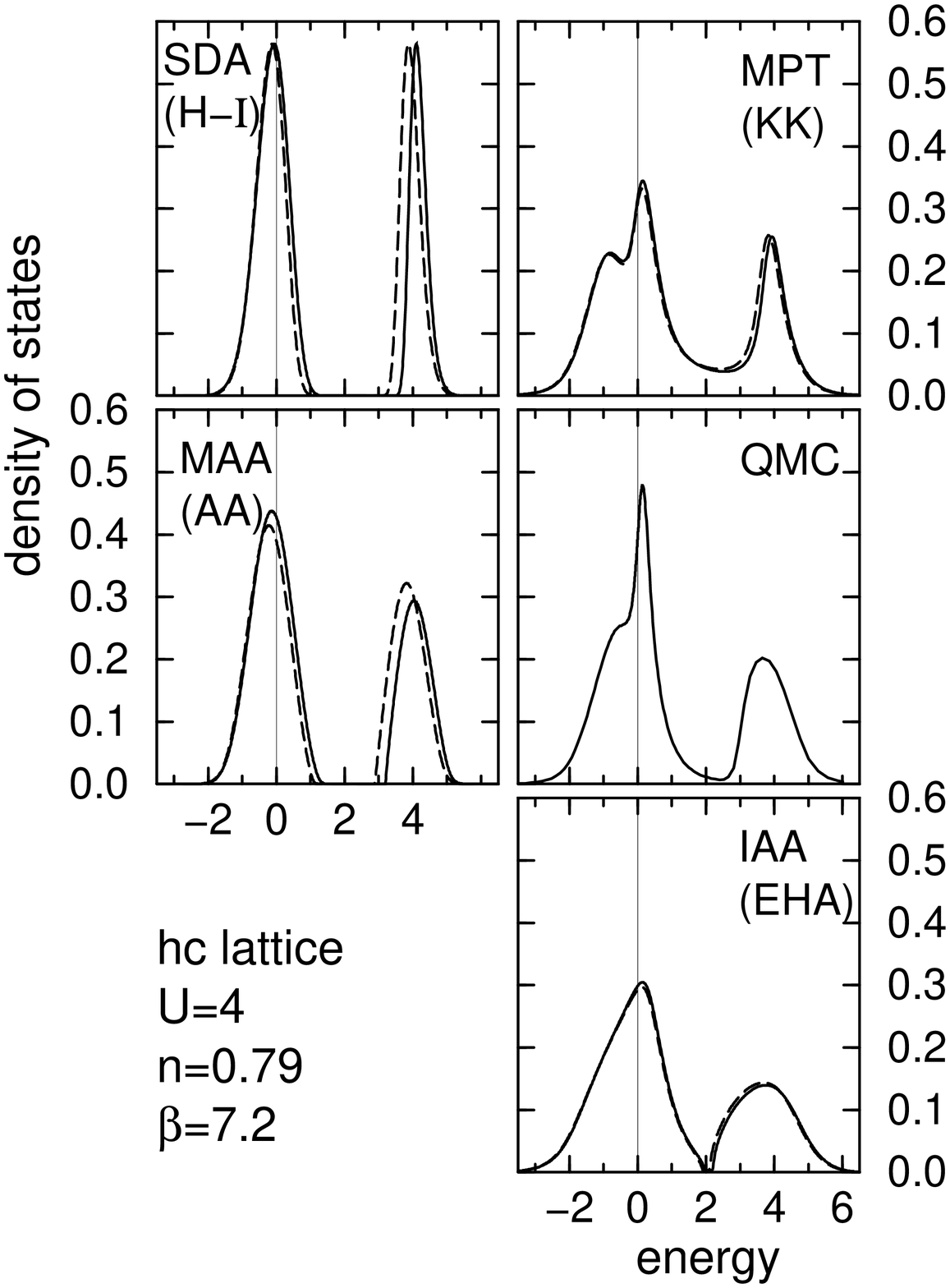,width=75mm,angle=0}}
\vspace{3mm}
\parbox[]{85mm}{\small Fig.~2.
Density of states $A_\sigma(E)$ as a function of energy for the hc 
lattice at $U=4$, $n=0.79$ and $\beta=7.2$ as obtained within the 
Hubbard-I solution (H-I), the conventional alloy analogy (AA), the 
Kajueter-Kotliar approach (KK) and the Edwards-Hertz approach (EHA):
dashed lines. Solid lines: resulting DOS within the improved theories: 
SDA, MAA, EHA and MPT. For comparison the numerically exact result
obtained within the Quantum Monte Carlo method (QMC) by Jarrell and
Pruschke \cite{JP93} is shown.
}
\end{figure}

\begin{figure}[t]
\vspace{8mm}
\centerline{ 
\psfig{figure=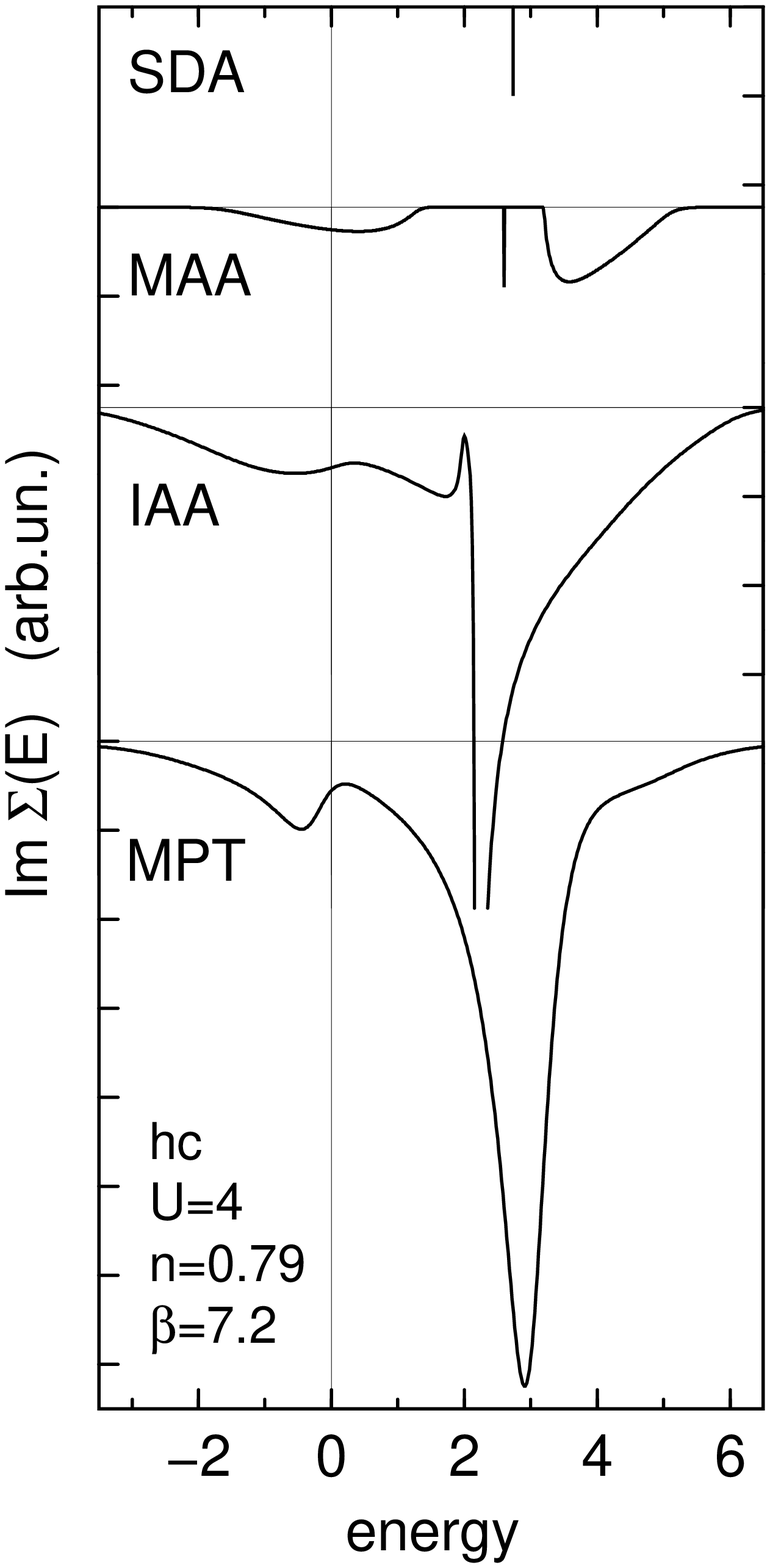,width=75mm,angle=0} \hspace{10mm}}
\vspace{4mm}
\parbox[]{85mm}{\small Fig.~3.
Imaginary part of the self-energy as a function of energy for the 
hc lattice at $U=4$, $n=0.79$ and $\beta=7.2$. Thin horizontal 
lines indicate $\mbox{Im} \, \Sigma(E) = 0$. Within the SDA 
$\mbox{Im} \, \Sigma(E)$ consists of a single delta peak only which 
is indicated by a vertical solid line at the calculated energetic 
position (also for the MAA). For the IAA $\mbox{Im} \, \Sigma(E)$ is
always finite, large (negative) values near $E=2.2$ are not shown.
}
\end{figure}

Comparing the other pairs of approximation schemes among each other,
we notice similar qualitative effects. Generally, the inclusion of 
$B_{-\sigma}$ in the theory results in fairly small changes of the 
lower but in a significant shift of the upper Hubbard band to 
higher energies (for fixed $n$). For the MAA there is an 
additional change in the peak heights compared with the AA.
For the IAA and the MPT the shift of the respective upper Hubbard
band against the EHA and the KK result turns out to be much 
smaller when compared with the MAA/AA or SDA/H-I. The reason is
that within the EHA and the KK approach the band shift is not
neglected completely (as for H-I and AA where 
$n_{-\sigma} B_{-\sigma} \mapsto n_{-\sigma} T_0$), but is taken
into account on the Hartree-Fock level at least:
$n_{-\sigma} B_{-\sigma} \mapsto n_{-\sigma} B_{-\sigma}^{\rm (HF)}$.
Indeed, as is shown in the second panel of Fig.~1, the difference
$n_{-\sigma} ( B_{-\sigma} - B_{-\sigma}^{\rm (HF)} )$ as 
calculated self-consistently within the IAA and the MPT is 
considerably smaller than the difference 
$n_{-\sigma} ( B_{-\sigma} - T_0 )$ relevant for the SDA and MAA.

Fig.~2 also shows the essentially exact result for the DOS as
obtained by Jarrell and Pruschke \cite{JP93} by means of the QMC
method. One clearly recognizes the two Hubbard bands which are
well separated from each other. The energetic difference is 
approximately given by $U$. Additionally, in the vicinity of 
$E=0$ there is a rather sharp quasi-particle resonance which is
reminiscent of the Kondo peak in the SIAM. All approximate methods
reliably reproduce the energetic positions as well as the weights
of the Hubbard bands. If at all, there is only a slight 
improvement with respect to positions and weights when taking
into account the correlation function $B_{-\sigma}$ additionally. 
We can 
conclude that in this respect and in the case of the paramagnet
the effects of $B_{-\sigma}$ are rather unimportant; considering
the moment sum rule does not help much to improve the agreement
with the QMC result.

Since quasi-particle damping is neglected completely within the SDA,
the Hubbard peaks turn out to be too narrow. Furthermore, there is
no additional structure at $E=0$. Obviously, the Kondo-type 
resonance cannot evolve in a two-pole approach. Somewhat broader
peaks are predicted by the MAA; compared with the QMC density of
states, however, their width is still too small. The IAA yields
a considerable improvement. Yet, the size of the Hubbard gap is
underestimated, and only a small asymmetry in the shape of the lower
Hubbard band hints towards the resonance. The MPT achieves a 
reasonable agreement with the exact result. The Kondo-type 
resonance is clearly visible but not as pronounced as in the 
QMC spectrum.

Additional insight is provided by the imaginary part of the
retarded self-energy which is shown in Fig.~3. The pole of the
SDA self-energy (\ref{eq:sdasig}) at $E=B_{-\sigma} + U ( 1 - 
n_{-\sigma}) - \mu$ gives rise to a single delta peak in 
$\mbox{Im} \, \Sigma(E)$, which, however, is meaningless since it
falls into the Hubbard gap. Taking the imaginary part of the 
CPA equation (\ref{eq:cpaeq}) at an energy $E$ where 
$\mbox{Im} \, G(E) = 0$ (is exponentially small), we obtain:
\begin{equation}
   0 = \mbox{Im} \, \Sigma_\sigma(E) \sum_{p=1}^2
   \frac{x_{p\sigma}}{\left| 1 - \frac{1}{\hbar} 
   G_\sigma(E) (E_{p\sigma} - \Sigma_\sigma(E) - T_0) \right|^2}
   \: .
\label{eq:imcpa}
\end{equation}
This implies that a finite imaginary part of the MAA self-energy
can be found in those energy ranges with non-vanishing DOS only,
except for the energy point at which there is a zero of the 
denominator in Eq.\ (\ref{eq:imcpa}). Here we find a delta peak
in $\mbox{Im} \, \Sigma(E)$ again (vertical line in Fig.~3). The
delta peak does not contribute to the damping of the DOS but 
acquires most of the weight. This explains why the broadening of the
Hubbard bands is still underestimated by the MAA. 

In the IAA $G_\sigma(E)$ is replaced by the modified propagator
$\widetilde{G}_\sigma(E)$ which yields a considerably stronger
quasi-particle damping. We also note a small dip in the imaginary
part of the IAA self-energy at $E \approx 0$. This tends to 
reproduce the correct Fermi-liquid behavior $\mbox{Im} \,
\Sigma(E) \sim E^2$ for $T=0$. A $T=0$ calculation, however, 
shows $\mbox{Im} \, \Sigma(0) \ne 0$ at $U=4$ and $n=0.79$.
Similar to the EHA \cite{EH,WC}, the IAA does not 
yield a Fermi surface in wide regions of the $U$-$n$ plane. 
Contrary, the MPT self-energy always vanishes quadratically
at $E=0$ and $T=0$. Fig.~3 also shows a more pronounced dip
in $\mbox{Im} \, \Sigma(E)$ at $E \approx 0$ in the
finite-temperature MPT result. Compared with the IAA, the
damping effect is of the same order of magnitude. In all cases 
the total weight $\int \mbox{Im} \, \Sigma_\sigma(E) \, dE$ is 
given by $-\pi U^2 n_{-\sigma} (1-n_{-\sigma})$. This follows
directly from the high-energy asymptotics of the self-energy
(\ref{eq:smom}).

\begin{figure}[b]
\vspace{4mm}
\centerline{\psfig{figure=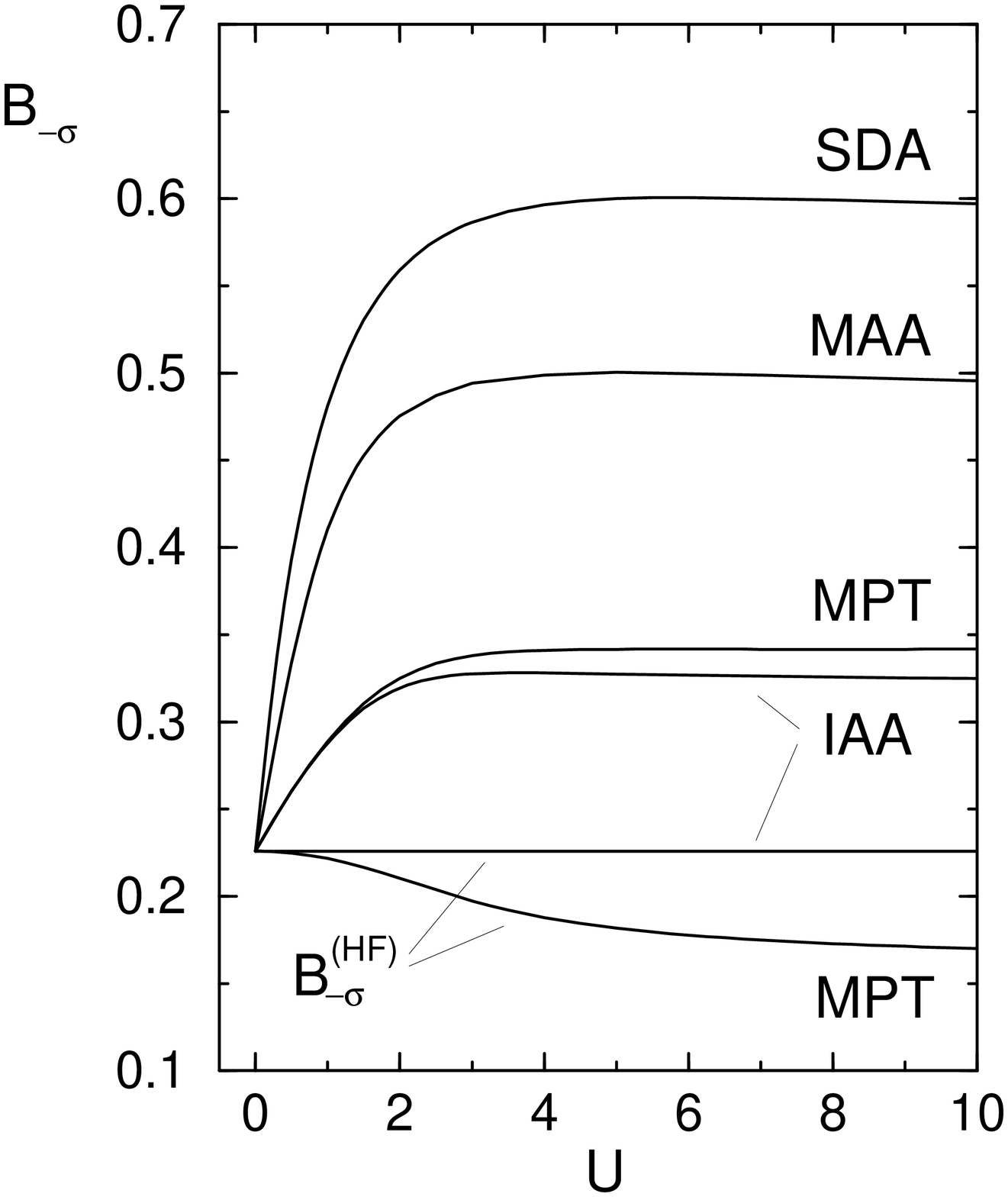,width=75mm,angle=0}}
\vspace{-6mm}
\parbox[]{85mm}{\small Fig.~4.
$U$ dependence of the correlation function $B_{-\sigma}$
(hc lattice, $n=0.79$, $\beta=7.2$). For the IAA and
the MPT $B_{-\sigma}^{\rm (HF)}$ is shown additionally.
}
\end{figure}

There is a strong dependence of the band shift on the interaction
strength $U$. Fig.~4 shows $B_{-\sigma}$ as a function of $U$ as
resulting from the different approximation schemes for $n=0.79$ 
and $\beta =7.2$. For large $U$ the correlation function 
$B_{-\sigma}$ becomes essentially independent of $U$. Within all
approaches this asymptotic behavior is reached for $U\approx 4$;
for $U > 4$ there is a very slight $U$ dependence only. 
The asymptotic values predicted by the SDA and the MAA are
considerably larger than those predicted by the IAA and MPT.
Although the underlying physical concepts are quite different,
there is a remarkable similarity between the results from 
the latter two approaches over the whole $U$ range. For small $U$ 
the results are nearly identical (up to $U\approx 2$). This had 
to be expected since the IAA as well as the MPT self-energies are 
correct up to order $U^2$. Via Eq.\ (\ref{eq:bfin}) this implies 
that $B_{-\sigma}(U)$ is exact in order $U$, i.~e.\ at $U=0$ 
$B_{-\sigma}(U)$ starts with the correct slope within the IAA 
and MPT. Of course, at $U=0$ all approaches yield the same value 
for $B_{-\sigma}$.

For the IAA and the MPT Fig.~4 also shows the dependence of 
$B_{-\sigma}^{\rm (HF)}$ on $U$. Within the IAA 
$B_{-\sigma}^{\rm (HF)}$ is given by Eq.\ (\ref{eq:b1def}).
The condition (\ref{eq:nnhf}) that fixes the shifts $E_\sigma$
can be written as $n^{(1)}_\sigma = n_\sigma$. It implies that 
$n^{(1)}_\sigma$ as well as 
$\langle c^\dagger_{i\sigma} c_{j\sigma} \rangle^{(1)}$ 
are independent of $U$, and therefore $B_{-\sigma}^{\rm (HF)}$ 
turns out to be a constant that depends of $n$ and $\beta$ only.
Analogously, due to the condition (\ref{eq:c2}) there is no
$U$ dependence of $n_\sigma^{(1)}$ in the MPT. The hybridization 
function $\Delta_\sigma$, however, does depend on $U$ due to the 
self-consistent mapping procedure. This leads to a $U$ dependence 
of the second factor in Eq.\ (\ref{eq:b1siam}) and thus of 
$B_{-\sigma}^{\rm (HF)}$ as can be seen in Fig.~4.

For all approaches considered here we did not find ferromagnetic
solutions on the hc lattice. We calculated the uniform static
susceptibility by applying an infinitesimally small external field:
$\chi=\partial m/\partial H$. The susceptibility never diverged for
$U < 6$. This is consistent with the QMC results of Refs.\ 
\cite{GKKR96,PJF95} and also with the complete instability of 
the Nagaoka state on the hc lattice \cite{Uhr96}. Let us mention,
however, that a partially polarized ferromagnetic state has been
found recently for extremely strong interaction $U$ \cite{OPK97}.
\\

{\center \bf \noindent B. Ferromagnetism \\ \mbox{} \\} 

While the moment sum rule for $m=3$ has turned out to be of marginal 
importance for the paramagnetic phase, it is much more decisive with
respect to the possibility and the characteristics of ferromagnetic
order. This shall be elucidated by the following arguments:

Firstly, ferromagnetism is a strong-coupling phenomenon. We thus
have to account for the Hubbard splitting. It is a necessary condition
for the evolution of the Hubbard bands as $U\mapsto \infty$, however, 
that the moment sum rule is obeyed up to $m=2$ (see Ref.\ \cite{VT97},
for example). This is achieved e.~g.\ by the H-I and the AA solution.

Secondly, within the H-I approach ferromagnetism turns out to be 
rather unlikely. Ferromagnetic order may be possible for small $n$ 
only when there is a large BDOS at the Fermi energy away from the 
band center of gravity \cite{Hub63}. The AA solution is even more 
prohibitive with respect to ferromagnetic order \cite{FE73,SD75}. The 
{\em reason} \cite{SD75} for this tendency to paramagnetism is that 
even in a (possible) ferromagnetic phase, the band centers of gravity 
$T_{p\sigma}({\bf k})$ of the majority and of the minority electrons 
at each ${\bf k}$ point are {\em equal}: $T_{p\uparrow}({\bf k}) = 
T_{p\downarrow}({\bf k})$ for both, the lower ($p=1$) and the upper 
($p=2$) Hubbard band. What is missing in the H-I and the AA approach 
to get a stable and extended ferromagnetic phase, is the possibility 
for a {\em spin-dependent} shift of the centers of gravity of the 
Hubbard bands.

Thirdly, the possibility for a spin-dependent band shift is not 
only required but in fact is predicted by the $1/U$ perturbational 
approach of Harris and Lange which is exact for $U\mapsto \infty$: 
As is seen from Eq.\ (\ref{eq:hl}), the spin-dependent shifts of 
the lower and the upper $\sigma$ band are $n_{-\sigma} B_{-\sigma}$ 
and $(1-n_{-\sigma}) B_{-\sigma}$, respectively. The higher-order 
correlation functions included in the definition (\ref{eq:bdef}) 
of $B_{-\sigma}$ open the channel for ferromagnetic solutions in 
the Hubbard model.

Finally, as has been discussed in Sec.~2, the perturbational results 
of Harris and Lange for the strong-coupling regime can only be 
recovered if the moment sum rule is respected up to $m=3$. 

It goes without saying that the $1/U$ perturbation theory merely
determines the global features of the quasi-particle spectrum
and abstracts from any internal structure of the Hubbard bands 
which may be due to damping effects or low-energy excitations, for 
example. Indeed, in the preceding section we have seen that different 
approximations yield rather different results even if all respect
the sum rule up to $m=3$. Thus it has to be expected that there is 
a dependence of the results on the approximation used also in the 
case of the ferromagnet. 

Our intention here is twofold:
Firstly, as has been noticed in the case of the paramagnet, 
conceptually improving the method (SDA $\mapsto$ MAA $\mapsto$ 
IAA, MPT) also yields a closer agreement with the QMC results. 
This will be checked for the ferromagnet, too. Secondly, comparing
the results of different methods with each other 
(H-I $\leftrightarrow$ SDA, AA $\leftrightarrow$ MAA,
EHA $\leftrightarrow$ IAA, KK $\leftrightarrow$ MPT), we are able to 
``switch on'' and to ``switch off'' in a controlled way the possibility 
for a spin-dependent band shift that is given via $B_{-\sigma}$.
This will clarify whether the concept of the spin-dependent band 
shift really helps to understand ferromagnetism, i.~e.\ whether 
the system ``uses'' this possibility and ``realizes'' the 
spin-dependent shift of the bands.

\begin{figure}[t]
\vspace{-8mm}
\centerline{\psfig{figure=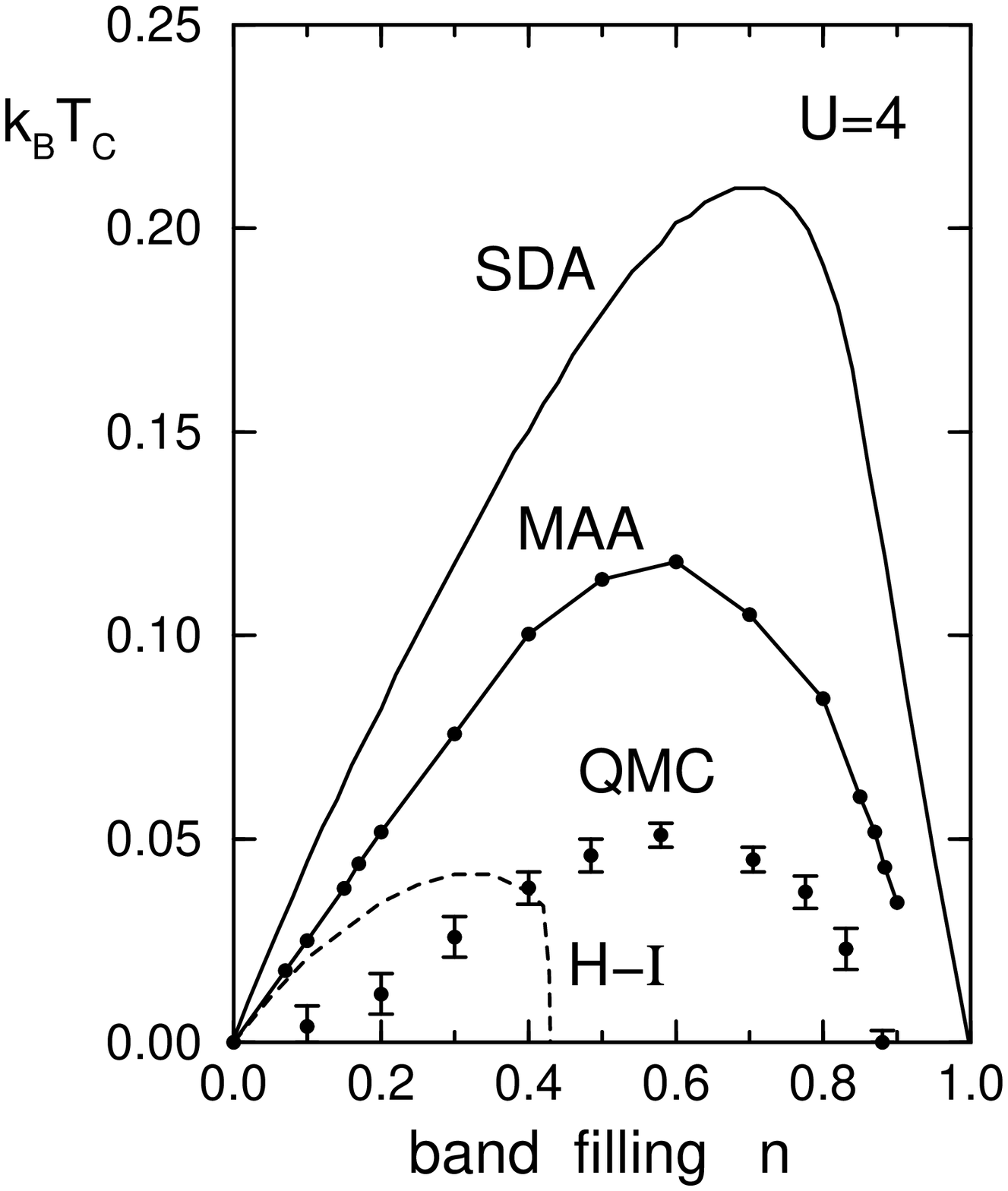,width=90mm,angle=0}}
\vspace{-23mm}
\parbox[]{85mm}{\small Fig.~5.
Filling dependence of the Curie temperature $T_C$ for the SDA 
(solid line) and H-I (dashed line) as well as for the MAA (circles) 
in comparison with the QMC results (error bars) by Ulmke \cite{Ulm98}
for the $d=\infty$ fcc-type lattice.
(Within the AA there is no ferromagnetic instability at all).
}
\end{figure}

\begin{figure}[t]
\vspace{-8mm}
\centerline{\psfig{figure=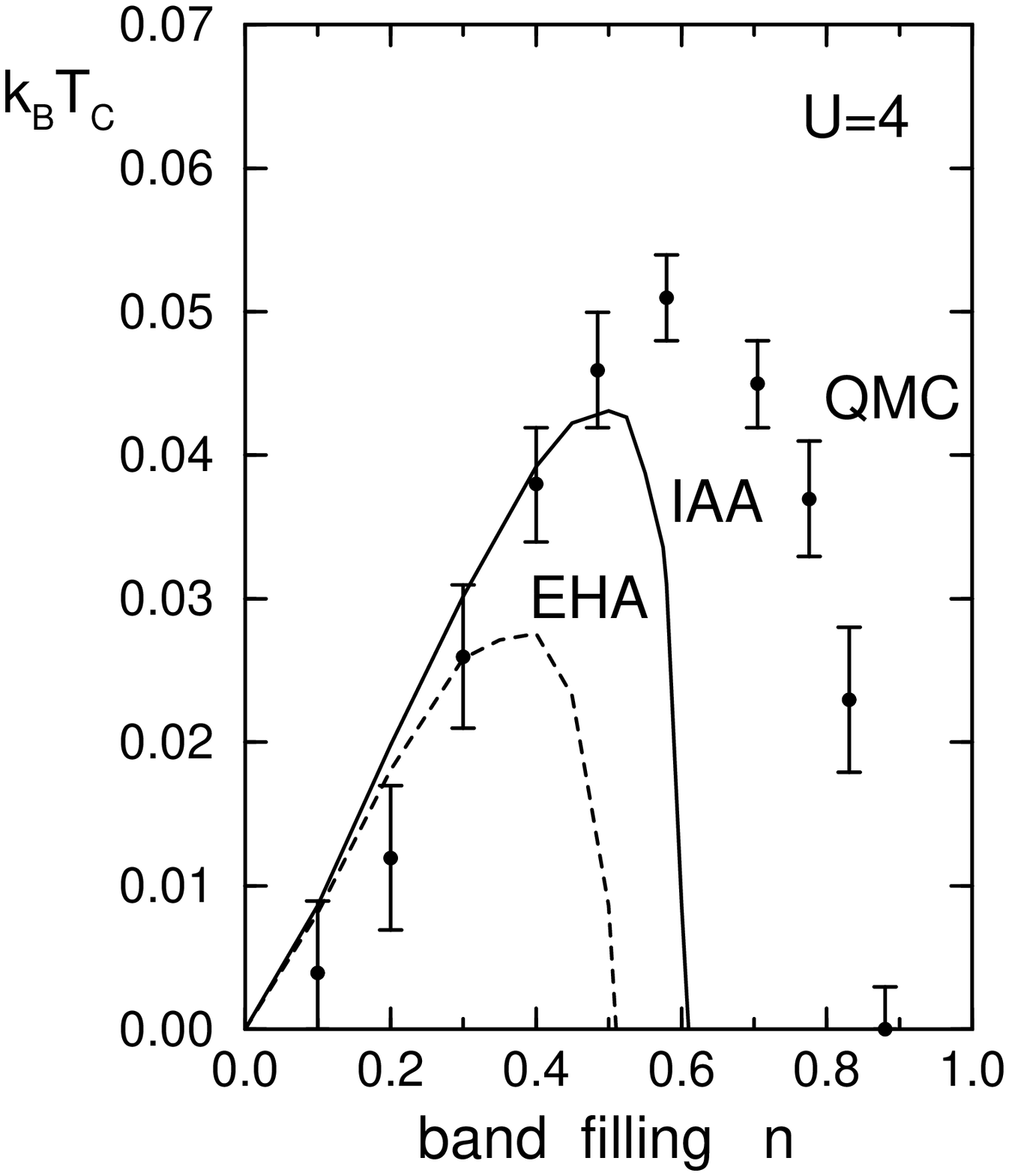,width=90mm,angle=0}}
\vspace{-23mm}
\parbox[]{85mm}{\small Fig.~6.
$T_C(n)$ as obtained within the IAA (solid line) and the EHA (dashed
line) compared with the QMC result \cite{Ulm98}.
}
\end{figure}

\begin{figure}[t]
\vspace{-8mm}
\centerline{\psfig{figure=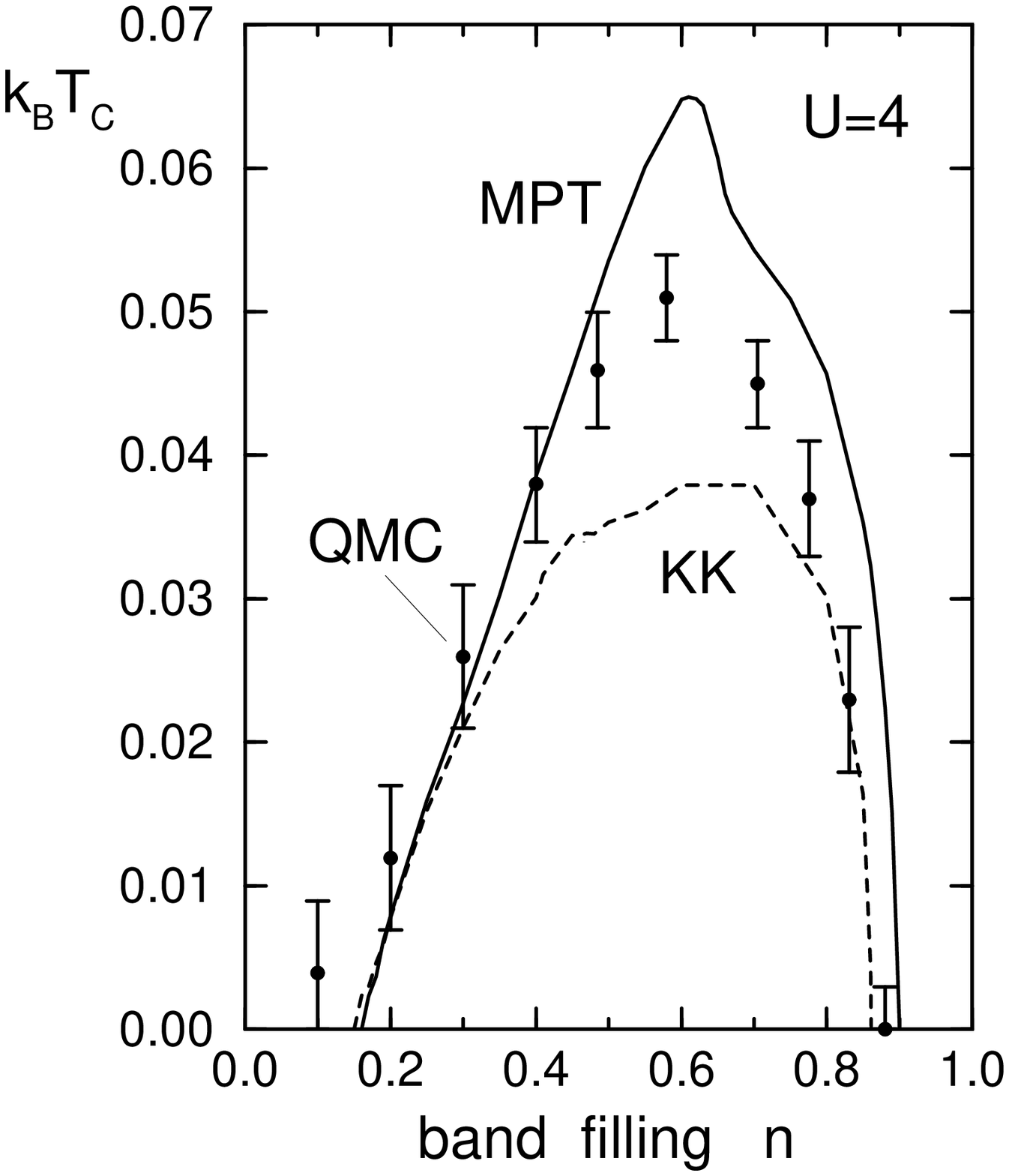,width=90mm,angle=0}}
\vspace{-23mm}
\parbox[]{85mm}{\small Fig.~7.
$T_C(n)$ as obtained within the MPT (solid line) and KK (dashed
line) compared with the QMC result \cite{Ulm98}.
}
\end{figure}

For the infinite-dimensional fcc-type lattice with the BDOS given
by Eq.\ (\ref{eq:bdosfcc}), ferromagnetic order even occurs at a 
moderate coupling strength $U=4$ and extends over a rather wide
region in the $n$-$T$ plane. This was proven numerically within the
QMC method by Ulmke \cite{Ulm98}. Due to the strong frustration of
the fcc-type lattice antiferromagnetic order is not expected. 
The highly asymmetric BDOS (\ref{eq:bdosfcc}) with its square-root 
singularity at the lower band edge $E_b=-1/\sqrt{2}$ favors 
ferromagnetic order \cite{VBH+97,Ulm98,WBS+97}. The QMC results 
of Ulmke \cite{Ulm98} are shown in Fig.~5. The filling-dependence 
of the Curie temperature $T_C$ has been obtained from the zero of 
$\chi^{-1}$ which shows a linear Curie-Weiss behavior.

The high density of states at the lower band edge is sufficient
to produce stable ferromagnetism within the H-I solution. The
resulting filling dependence of $T_C$ is shown in Fig.~5. The 
ferromagnetic region in the $n$-$T$ plane, however, is confined
to very low densities only. As mentioned this is known to be 
typical for the H-I approach \cite{Hub63}. Contrary to the H-I 
solution, the SDA allows for a spin-dependent band shift. Fig.~5 
shows that this leads to a considerable increase of the Curie 
temperature. Furthermore, ferromagnetic solutions are found for 
all fillings up to half-filling $n=1$. These effects of $B_{-\sigma}$ 
are much stronger than those found for the paramagnetic phase. 

At $n=0.58$ and $T=0$ the system is fully polarized ($m=n$). The 
spin splitting of the center of gravity of the lower Hubbard band 
amounts to $n_\uparrow B_\uparrow - n_\downarrow B_\downarrow = 0.73$.
This has to be removed as $T$ approaches the Curie temperature while 
the Hubbard gap still exists above $T_C$. The spin splitting is 
considerably smaller compared with the exchange splitting 
$U n_\uparrow - U n_\downarrow = 2.32$ within Hartree-Fock (Stoner) 
theory where correlation effects are neglected altogether. 
Consequently, the SDA yields a Curie temperature that is 
smaller by more than a 
factor 3: $k_B T_C^{\rm (SDA)} = 0.19$ and $k_B T_C^{\rm (Stoner)} = 
0.74$ (at $n=0.58$, $U=4$). The difference becomes larger and 
larger with increasing $U$ since as a function of $U$ the Curie
temperature is unbounded in the Stoner theory while it reaches 
saturation within the SDA \cite{HN97b}. The unrealistically high 
Stoner Curie temperature results from the necessity to bridge an 
exchange splitting of order $U$ by the thermal energy $k_B T_C$. 

A spin-dependent band shift is also realized within the MAA 
approach. Stable ferromagnetism is found for fillings $0<n<1$ 
(close to half filling we have not succeeded to obtain truely 
converged numerical solutions). Compared with the SDA, the Curie
temperature is significantly lower for all fillings. While the
mechanism leading to ferromagnetic symmetry breaking is the same 
in the SDA and MAA, ferromagnetism is destabilized to some extent 
by means of quasi-particle damping (see also Ref.\ \cite{HN96}). The 
broadening of the respective spin-dependent bands tends to enhance 
their overlap and leads to a (self-consistent) depression of the 
magnetization and thereby of $T_C$.
The decisive meaning of the $m=3$ sum rule becomes most apparent
when we ``switch off'' the possibility of a spin-dependent band 
shift, i.~e.\ if we compare the MAA results with those of the 
conventional AA. There is no ferromagnetic instability at all in the
AA as has been proven in Ref.\ \cite{FE73} and also tested numerically
for the fcc-type lattice considered here. Therefore, ignoring the 
possible spin-dependent band shift $n_{-\sigma}B_{-\sigma}$, misses
the most important route to ferromagnetic order.

Quasi-particle damping becomes stronger when turning from the MAA 
to the IAA. Again, we notice a considerable decrease of the Curie
temperature for all fillings. Fig.~6 shows that the maximum $T_C$
in the IAA is less than half of the maximum $T_C$ predicted by the
MAA. Moreover, the ferromagnetic region shrinks to $n<0.61$. 
Comparing with the EHA, we again notice that the spin-dependent band 
shift induced by $B_{-\sigma}$ favors ferromagnetic order: Since 
$B_{-\sigma}$ is replaced by $B_{-\sigma}^{\rm (HF)}$ in the EHA 
[cf.\ Eq.\ (\ref{eq:ehahigh})], the EHA partially accounts for the
band shift. However, we find the spin splitting of $B_{-\sigma}$ 
to be larger than that of $B_{-\sigma}^{\rm (HF)}$. Consequently,
$T_C$ is enhanced in the IAA although the differences seen in 
Fig.~6 are moderate compared with the cases H-I/SDA and AA/MAA.

The same arguments hold for the MPT and the KK approach: The spin 
splitting of the center of gravity due to $B_{-\sigma}^{\rm (HF)}$
is somewhat smaller than that due to $B_{-\sigma}$. This implies
a higher $T_C$ in the MPT compared with the KK approach [see Eq.\ 
(\ref{eq:kkhigh})]. The filling dependence of $T_C$ is shown in
Fig.~7. The differences between the KK approach and the MPT are 
moderate again and comparable to the EHA/IAA case. Once more, this 
suggests a one-to-one correspondence between the critical temperature 
and the band shift due to $B_{-\sigma}$ in the $m=3$ moment.

For comparison we included the essentially exact QMC result 
\cite{Ulm98} for $T_C(n)$ in Figs.~5-7. While the Stoner theory 
yields a Curie temperature that is by more than one order of 
magnitude too high (see Ref.\ \cite{WBS+97}, for example), 
realistic values are found by the approximations discussed 
here. Even the simplest scheme consistent with the requirements
of Harris and Lange, the SDA, considerably improves upon the 
Stoner result. A quantitative agreement with the essentially 
exact result, however, cannot be expected since damping effects 
are neglected completely. A further improvement towards
the QMC result is achieved by the MAA which takes into account 
a finite quasi-particle life time. Yet, the Curie temperatures are 
systematically too high since damping effects are underestimated 
(cf.\ previous section and Fig.~9). The correct order of magnitude 
for $T_C$ can only be expected if (i) the theory respects the 
$m=3$ sum rule and (ii) realistically accounts for quasi-particle 
damping as is done by the IAA and the MPT.

Fig.~8 focuses on the temperature dependence of the magnetization 
at $n=0.58$. The QMC data can be well fitted to an $S=1/2$ 
Brillouin-function. Extrapolation to $T=0$ yields $m=n$ \cite{Ulm98}. 
A fully polarized ground state is also obtained by the SDA and MAA, 
the MPT predicts a slightly lower $m$. The filling $n=0.58$ is near 
to the quantum-critical point $n_c^{\rm (IAA)}=0.61$ in the IAA 
(cf.\ Fig.~6). Since phase transitions 
have been found to be always of second order within the IAA, a 
partially polarized ground state near $n_c^{\rm (IAA)}$ is plausible  
from the result in Fig.~6. The vicinity to $n_c^{\rm (IAA)}=0.61$ 
may also explain the non-Brillouin-function-like shape of the 
magnetization curve. Fig.~8 shows second-order transitions for 
the SDA, MAA and MPT at $n=0.58$. Depending on the filling,
however, all three methods may also predict first-order transitions 
to the paramagnetic phase. Non-continuous transitions are observed 
for fillings larger than $\sim n_{\rm max}$ where $n_{\rm max}$ is 
the filling at which $T_C$ reaches its maximum. For the MPT the 
change of second-order into first-order transitions is marked by 
the kink in the $T_C(n)$ curve at $n=0.67$ (Fig.~7). While magnetic
first-order transitions may seem to be implausible, they cannot be
ruled out by a rigorous argument. On the other hand, we cannot 
strictly exclude that they are artefacts of the approximations.

\begin{figure}[tbp]
\vspace{6mm}
\centerline{\psfig{figure=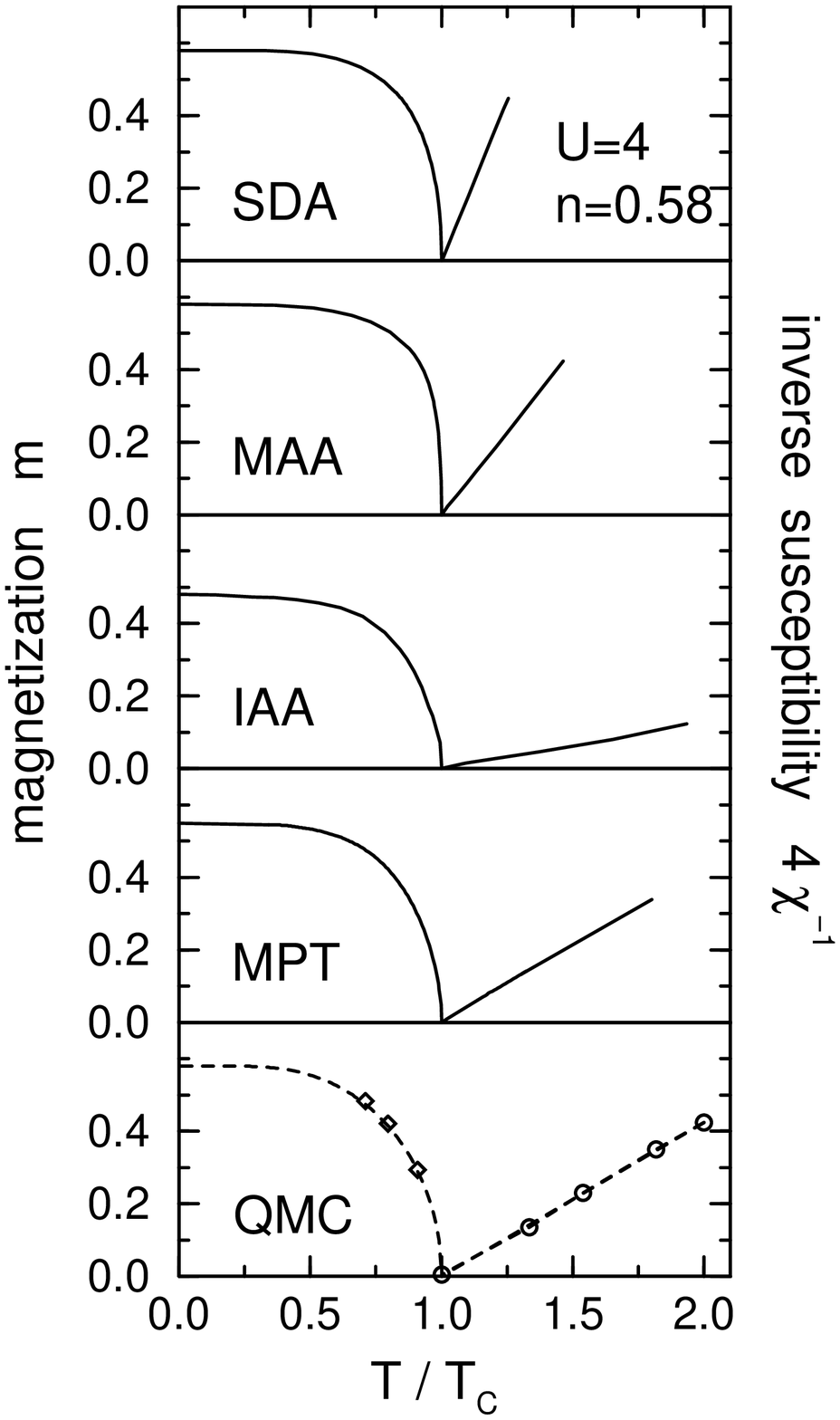,width=100mm,angle=0}}
\vspace{-12mm}
\parbox[]{85mm}{\small Fig.~8.
Homogeneous magnetization $m$ as a function of the reduced temperature
$T/T_C$ for the fcc-type lattice at $U=4$ and $n=0.58$
as obtained self-consistently within the SDA, MAA, IAA and the MPT
(solid lines below $T/T_C=1.0$). Bottom panel: QMC data from
Ulmke \cite{Ulm98} (diamonds). The dashed line below $T/T_C=1.0$
is a Brillouin-function fit to the data. For $T/T_C>1.0$ the inverse 
homogeneous static susceptibility $\chi^{-1}$ (multiplied by a factor 
4) is shown. Circles: QMC data \cite{Ulm98}. Dashed line: linear fit 
to the data.
}
\end{figure}

\begin{figure}[tbp]
\vspace{4mm}
\centerline{\psfig{figure=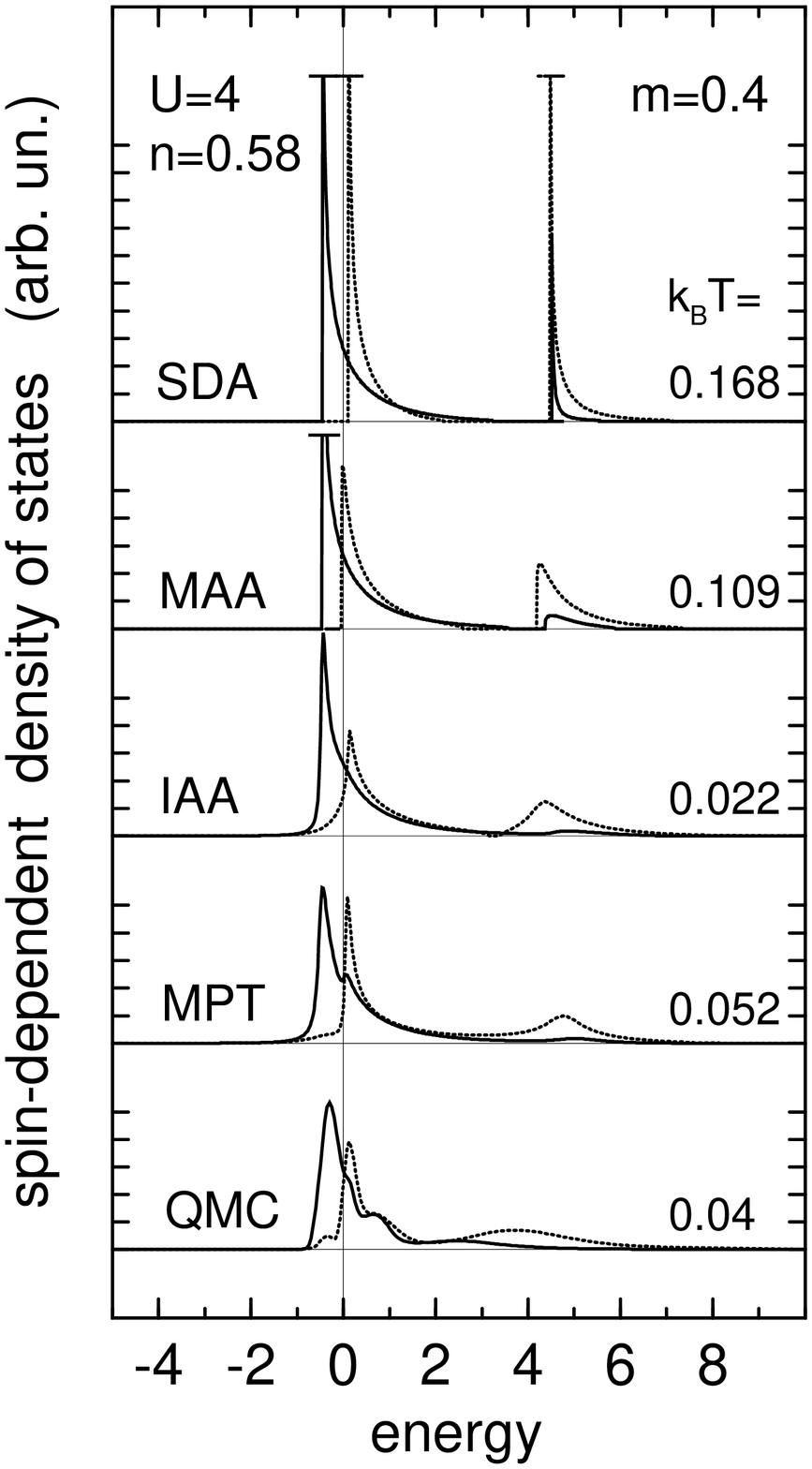,width=85mm,angle=0}}
\vspace{3mm}
\parbox[]{85mm}{\small Fig.~9.
Spin-dependent density of states $A_\sigma(E)$ as a function of 
energy (fcc lattice, $U=4$, $n=0.58$). Solid (dotted) lines: 
majority (minority) spin. QMC result from Ref.\ \cite{Ulm98}.
In each case (SDA, MAA, IAA, MPT) the respective temperature 
(given in the figure) has been chosen to yield a magnetization 
of $m=0.4$.
}
\end{figure}

For the second-order transitions shown in Fig.~8 the homogeneous,
static susceptibility $\chi=\partial m/\partial H |_{H=0}$ diverges
at $T=T_C$. Within all approaches considered $\chi$ obeys the 
Curie-Weiss law for high temperatures. But also for temperatures 
close to $T_C$ we observe an almost linear trend of $\chi^{-1}$. 
This is consistent with the QMC result. Note that the different 
slopes in the plot result from the different Curie temperatures.
From $\chi = C / ( k_{\rm B} T - k_{\rm B} \Theta )$ we have
calculated the (dimensionless) Curie constant $C$. For the 
paramagnetic Curie temperature we have assumed $\Theta = T_C$ 
except for the IAA where $k_{\rm B} \Theta = 0.057$ has been 
obtained by extrapolation of the linear trend of $\chi^{-1}$ at high
temperatures. The QMC value $C=0.47$ is to be compared with 
$C=0.42, 0.52, 0.50, 0.57$ for the SDA, MAA, IAA and MPT, respectively.

Finally, we compare the spin-dependent spectral density as obtained
by the approximate schemes with the essentially exact QMC result.
Fig.~9 shows $A_\sigma(E)$ in the ferromagnetic phase for $\sigma = 
\uparrow$ (majority) and $\sigma = \downarrow$ (minority spin) at
$U=4$ and $n=0.58$. For a meaningful comparison the respective 
temperature has been chosen to yield a magnetization of $m=0.4$.
In the SDA spectrum we recognize the large Hubbard splitting $\sim U$
and a much smaller additional spin splitting for each of the two
Hubbard bands. Furthermore, a correlation-induced spin-dependent 
narrowing as well as a spin-dependent weight of the subbands is 
observed. All features agree qualitatively well with the predictions 
of Harris and Lange which are valid for $U\mapsto \infty$. A more 
quantitative estimate shows up discrepancies with respect to the 
positions and weights given by Eq.\ (\ref{eq:hl}) which are due to the 
moderate value assumed for $U$. 

The shape of the peaks only slightly 
deviates from the shape of the BDOS. This is an artefact resulting from 
the complete disregard of quasi-particle damping. Turning to the MAA 
spectrum, a broadening of the peaks is introduced. Damping effects 
are even more pronounced by the IAA and the MPT. This turns out to 
be quite realistic when comparing with the QMC spectrum which also 
shows up a considerable peak broadening. In all approximations the 
spin-dependent position and weight of the lower Hubbard band 
is well reproduced, the position of the upper Hubbard band, however, 
is shifted to higher energies compared with the QMC spectrum. This 
indicates that for $U=4$ the interaction is not sufficiently strong
to reproduce the position resulting from the $1/U$ perturbation
theory. An additional peak in the $\sigma=\uparrow$ as well as in the 
$\sigma=\downarrow$ channel of the QMC spectrum shows up at 
$E\approx 0.8$ the physical origin of which is unclear. It is 
not reproduced by the approximate methods.
At $E\approx 0$ there is also a weak shoulder in the $\sigma=\uparrow$ 
QMC spectrum. This is clearly reproduced by the MPT; but also in the
IAA spectrum there is a (very weak) structure at $E\approx 0$. As for
the paramagnet the peak is interpreted as a Kondo-like resonance. 
It cannot show up in the $\downarrow$ spectrum, since the filling 
of the $\downarrow$ band is too small.
\\

{\center \bf \noindent VIII. CONCLUSIONS \\ \mbox{} \\} 

The present paper has discussed the moment sum rule as a valuable
source of a priori information on the physical properties of strongly
interacting lattice fermion models. The analysis has been restricted
to the single-band Hubbard model in infinite dimensions since this
allows to compare with numerically exact solutions. Here the
reliability of approximate approaches as well as the usefulness of 
the sum rule for practical improvements can be estimated safely.
The general trends that show up for $d=\infty$ are expected to hold
also for finite dimensions, where extensive use of the sum rule has
been made in the past.
Let us briefly list up the conclusions evolving from our analysis:

(i) The moment sum rule fixes the norm, the center of gravity, the
variance, etc.\ of the interacting spectral density at each ${\bf k}$
point in the Brillouin zone. To some extent this determines the
``global'' single-particle excitation spectrum for arbitrary $U$.
The sum rule is of special conceptual importance in the strong-coupling
regime. We could argue that any approximation that predicts the
Hubbard splitting for $U\mapsto \infty$ and that yields the correct
moments up to $m=3$ is fully consistent with the first non-trivial
results of the perturbation theory in $1/U$ by Harris and Lange. 
This includes important correlation effects such as the $U$, filling
and spin dependence of the centers of gravity, the widths and the
weights of the two Hubbard bands.

(ii) The high-energy expansion of the Green function and of the 
self-energy provides a practicable way to check to which order $m$
a particular approximation is consistent with the sum rule. The
majority of analytical but approximate approaches commonly used
turn out to be at variance with the sum rule for $m=3$. We 
considered four methods in detail: The Hubbard-I and the Hubbard-III
alloy-analogy solutions as well as the Edwards-Hertz and the 
Kajueter-Kotliar approaches. In each case the sum rule is fulfilled
up to $m=2$ only.

(iii) We have shown that the compatibility with the $m=3$ moment
sum rule can be restored by slight modifications of the original
methods. Starting from the H-I and the conventional AA solution,
this yields an optimized two-pole and an optimized alloy-analogy
approach which turn out to be identical with the spectral-density
approach and the so-called modified alloy-analogy solution, 
respectively. Analogously, the interpolating alloy-analogy-based 
approximation and the modified perturbation theory evolve 
straightforwardly from the EHA and the KK approach. We have checked 
that the applied modifications do not affect the validity of the 
original methods in exactly solvable limiting cases of the Hubbard 
model. 

(iv) All approximation schemes have been implemented for numerical
evaluation. We have presented and discussed new results for correlation
functions and dynamical quantities which are directly compared with
essentially exact QMC results on the hyper-cubic and an fcc-type
lattice in infinite dimensions. We observed that with increasing
complexity (which is necessary to recover an increasing number of 
exact limits) the agreement with the QMC results is improved. This
corroborates the usefulness of an interpolative analytical approach 
to the Hubbard model.

In particular, we found it necessary to account for damping effects
in a realistic way: If most of the total weight of the imaginary part
of the self-energy does not contribute to quasi-particle damping,
as in the MAA and more extremely in the SDA, the peak widths that
are obtained by QMC cannot be recovered. Comparing the results of 
the different methods we also found that quasi-particle damping has
a destabilizing effect on ferromagnetism which manifests itself in 
an overall decrease of the Curie temperature.

(v) Out of all approximation schemes considered here, the modified 
perturbation theory turns out to be most reliable. In the paramagnetic 
phase it correctly recovers the Kondo-type resonance, shows up the 
expected Fermi-liquid properties and approximately fulfills the 
Luttinger sum rule. The MPT yields the qualitatively correct density 
of states in the weak- and intermediate-coupling regime as confirmed 
by comparison with the QMC results. At half filling the approach 
reduces to the iterative perturbation theory (IPT) which is known to 
yield a rather realistic description of the Mott transition. For the 
ferromagnetic phase the MPT is able to predict the Curie temperature 
within an accuracy of $\sim 30\%$. Thereby, the MPT turns out to be 
superior compared with the IAA. The main drawback of the latter surely 
consists in its inability to predict a Fermi surface in wide regions 
of the phase diagram. The same defect is found in the MAA. However, 
the straightforward and physically well motivated concept of an 
``optimized alloy-analogy'' remains rather attractive. Finally, the
SDA is surely too simple to compete with the more elaborate methods.
On the other hand, its simplicity is advantageous when a quick though
rough estimation of the magnetic phase diagram is required.

(vi) The comparative opposition of the different methods has 
particularly shown the importance of the $m=3$ moment sum rule 
with respect to ferromagnetism. While the differences between 
the methods with and without regard of $B_\sigma$ are unimportant 
or even negligibly small for the paramagnetic phase, the correlation 
function $B_\sigma$ considerably affects the ferromagnetic/paramagnetic
phase boundary as well as the Curie temperature. The effect is most 
striking when comparing the AA, where ferromagnetic instabilities 
are ruled out completely, with the MAA which yields a reasonable 
dependence $T_C(n)$ for the fcc-type lattice and even 
overestimates $T_C$. The reason is found in the term $B_\sigma$ 
that appears first in the expression for the $m=3$ moment. It 
opens the possibility for a {\em spin-dependent} shift of the 
centers of gravity of the Hubbard bands. In the AA and also in 
the H-I solution this is missing completely ($B_\sigma \mapsto 
T_0$). In the EHA and the KK approach it is only approximately
accounted for ($B_\sigma \mapsto B^{\rm (HF)}_\sigma$). Since
$B^{\rm (HF)}_\sigma$ is found to be too small in the EHA and 
the KK approach, both methods yield a lowered $T_C$. 

From the recent QMC studies \cite{JP93,VBH+97,Ulm98}, in particular
from Ref.\ \cite{WBS+97}, but also from variational results
\cite{FMMH90,Uhr96,HUMH97}, the importance of the lattice structure
for ferromagnetism well established. It is striking that all 
presently considered approximations that account for $B_\sigma$
yield a ferromagnetic instability for the non-bipartite fcc-type 
lattice while (at least up to $U < 6$) the paramagnet is stable
on the hc lattice.
In Ref.\ \cite{VBH+97} Vollhardt et al.\ argue that a strongly
asymmetric BDOS with high spectral weight at the lower band edge
favors ferromagnetism because this minimizes the kinetic energy 
of the fully polarized state. The shape of the non-interacting 
BDOS is relevant since the $\uparrow$ DOS is unrenormalized if 
$n_{\uparrow}=n<1$. If it is assumed that details of the BDOS 
are not so relevant for the paramagnetic state where correlation
effects dominate, the ferromagnetic state becomes energetically 
more favorable. Considering the moment sum rule or, equivalently, 
the Harris and Lange results, the argumentation can be refined:
For the fully polarized state the expression (\ref{eq:bdef}) for 
the band shift reduces to $(1-n) n B_\uparrow = - \langle \sum T_{ij} 
c_{i\uparrow}^\dagger c_{j\uparrow} \rangle = - E_{\rm kin}$. This 
implies that a strong asymmetry of the BDOS not only minimizes the 
kinetic energy (at a given filling $n$) but also maximizes the shift 
$n_{\uparrow} B_{\uparrow} >0$ of the center of gravity of the lower 
$\downarrow$ Hubbard band to higher energies. Therewith, the spin 
splitting $n_{\uparrow} B_{\uparrow} - n_{\downarrow} B_{\downarrow} 
= n_{\uparrow} B_{\uparrow}$ of the lower Hubbard band becomes 
maximal which enhances the stability (reduces the instability) 
of the fully polarized state.
This mechanism is already present within the simple SDA. While the
approach overestimates the possibility for ferromagnetism, it 
predicts the correct trend when comparing different lattices
\cite{HN97a}.

As an intermediate to strong-coupling phenomenon, ferromagnetism
in all its different aspects is hardly amenable to a simple
explanation. We feel, however, that by considering the spin-dependent 
band shift appearing in $1/U$ perturbation theory and the $m=3$ moment 
one is able to capture some of the essentials in a rather simple and 
physically intuitive way.
\\

{\center \bf \noindent ACKNOWLEDGEMENT \\ \mbox{} \\} 

Financial support of this work by the Deutsche Forschungsgemeinschaft 
within the Sonderforschungsbereich 290 is gratefully acknowledged.
\\

\small
\baselineskip3.4mm

----------------------------------------------------------------------

\end{document}